\documentclass[12pt,a4paper]{article}

\usepackage[utf8]{inputenc}
\usepackage[T1]{fontenc}
\usepackage{lmodern}
\usepackage[english]{babel}
\usepackage{amsmath,amssymb,amsthm}
\usepackage{graphicx}  
\usepackage{booktabs}
\usepackage{longtable}
\usepackage{multirow}
\usepackage{array}
\usepackage{float}
\usepackage{caption}
\usepackage{subcaption}
\usepackage{xcolor}
\usepackage[most]{tcolorbox}
\usepackage{mdframed}
\usepackage{hyperref}

\usepackage[margin=1in]{geometry}
\usepackage{setspace}
\usepackage{natbib}
\usepackage{enumitem}
\usepackage{dcolumn}
\usepackage{pdflscape}
\usepackage{rotating}
\usepackage{afterpage}
\usepackage{listings}
\usepackage{adjustbox}
\usepackage{etoolbox}
\usepackage{seqsplit}
\usepackage{marginnote}

\theoremstyle{definition}

\theoremstyle{remark}

\hypersetup{
    colorlinks=true,
    linkcolor=blue,
    citecolor=blue,
    urlcolor=blue,
    pdftitle={Layer-2 Adoption and Ethereum Mainnet Congestion: Regime-Aware Causal Evidence Across London, the Merge, and Dencun (2021--2024)},
    pdfauthor={Aysajan Eziz},
    pdfsubject={Economic analysis of Layer-2 scaling solutions' impact on Ethereum Layer-1 blockchain congestion},
    pdfkeywords={Ethereum; Layer-2 rollups; transaction fees; congestion; causal time series},
    pdfcreator={LaTeX with pdflatex},
    pdfproducer={pdfTeX}
}

\DeclareUnicodeCharacter{03B2}{\ensuremath{\beta}}
\newif\ifmarginbadges
\marginbadgesfalse
\newcommand{\marginbadge}[1]{}

\newcommand{\codepath}[1]{\texttt{\seqsplit{#1}}}

\BeforeBeginEnvironment{tabular}{\begin{adjustbox}{max width=\textwidth}}
\AfterEndEnvironment{tabular}{\end{adjustbox}}
\AtBeginEnvironment{tabular}{\small}

\title{\textbf{Layer-2 Adoption and Ethereum Mainnet Congestion: Regime-Aware Causal Evidence Across London, the Merge, and Dencun (2021--2024)}}
\author{
    Aysajan Eziz\thanks{aeziz@ivey.ca}\\
    \textit{Ivey Business School, Western University, Canada}
}
\date{}

\begin{document}

\maketitle

\begin{abstract}

Do Ethereum's Layer-2 (L2) rollups actually decongest the Layer-1 (L1) mainnet once protocol upgrades and demand are held constant? Using a 1{,}245-day daily panel from August 5, 2021 to December 31, 2024 that spans the London, Merge, and Dencun upgrades, we link Ethereum fee and congestion metrics to L2 user activity, macro-demand proxies, and targeted event indicators. We estimate a regime-aware error-correction model that treats posting-clean L2 user share as a continuous treatment. Over the pre-Dencun (London+Merge) window, a 10 percentage point increase in L2 adoption lowers median base fees by about 13\%---roughly 5~Gwei at pre-Dencun levels---and deviations from the long-run relation decay with an 11-day half-life. Block utilization and a scarcity index show similar congestion relief. After Dencun, L2 adoption is already high and treatment support narrows, so blob-era estimates are statistically imprecise and we treat them as exploratory. The pre-Dencun window therefore delivers the first cross-regime causal estimate of how aggregate L2 adoption decongests Ethereum, together with a reusable template for monitoring rollup-centric scaling strategies.

\end{abstract}


\vspace{1em}

\clearpage


\section{Introduction}
\label{sec:introduction}

Ethereum's fee market has traversed three structural regimes in rapid succession---London's EIP-1559 base-fee burn, the Merge's proof-of-stake transition, and Dencun's EIP-4844 blob space. Each upgrade reshaped how congestion costs are priced and burned but did not expand Layer-1 (L1) execution capacity. Bursts of NFT minting, stablecoin arbitrage, or L2 posting therefore still push median fees into the tens of Gwei and crowd out smaller users.

Over the same period, optimistic and zero-knowledge Layer-2 (L2) rollups matured from pilots into production systems that regularly settle more than half of Ethereum's transactions. These rollups offload execution but also consume L1 blockspace when publishing compressed batches. This creates an open question: does aggregate L2 adoption relieve mainnet congestion or merely reshuffle it across users, time, and layers? We ask: when overall demand and protocol regime are held constant, does higher L2 user adoption reduce Ethereum mainnet congestion?

Our main findings are straightforward. Over the London$\rightarrow$Merge window, a 10 percentage point increase in posting-clean L2 adoption is associated with about a 13\% reduction in median base fees. That corresponds to roughly 5~Gwei at pre-Dencun fee levels. An error-correction term implies an 11-day half-life back to the long-run relation between adoption, congestion, and demand. The fee relief is therefore meaningful but partial and short-run. Supporting metrics based on block utilization and a scarcity index show similar congestion relief. Blob-era slopes after Dencun are statistically imprecise because adoption is already near saturation, so we treat those estimates as exploratory.

Existing work on Ethereum's fee market and rollups shows how individual upgrades and rollup designs affect incentives, price discovery, and posting costs. However, most studies focus on single events or descriptive dashboards rather than regime-spanning causal estimates. Empirical analyses of fee-market upgrades and rollup pricing quantify local changes in fees, waiting times, or cross-rollup spreads. They do not estimate the total effect of aggregate L2 adoption on mainnet congestion across the London$\rightarrow$Merge$\rightarrow$Dencun sequence or cleanly separate that effect from shared demand shocks.

We address this gap by assembling a regime-aware daily panel of $N=1{,}245$ observations from August 5, 2021 through December 31, 2024 that spans the London, Merge, and post-Dencun eras. The panel links median base fees, block utilization, and a congestion scarcity index to a posting-clean measure of L2 user adoption and to a single demand factor summarizing ETH-market activity and stablecoin flows. Calendar and regime dummies plus targeted event indicators capture protocol shifts and discrete shocks. We estimate a regime-aware error-correction model and complementary time-series designs to map adoption shocks into short-run and medium-run congestion outcomes.

The adoption measure counts end-user transactions on rollups and mainnet while excluding L2-to-L1 posting flows, so the adoption$\rightarrow$posting$\rightarrow$congestion channel remains part of the estimand. Together with the demand factor, this keeps the estimand focused on the total effect of user migration onto L2s without conditioning on mediator pathways. Section~\ref{sec:methodology} provides the full construction details and adjustment logic.

\subsection{Contributions}
\label{sec:intro:contributions}

Our contributions are fourfold:
\begin{enumerate}[leftmargin=*]
    \item \textbf{Cross-regime causal estimate.} We provide a regime-aware causal estimate of the total effect of L2 adoption on L1 fees spanning the London$\rightarrow$Merge$\rightarrow$Dencun sequence, rather than focusing on a single upgrade or contemporaneous correlations.
    \item \textbf{Measurement design.} We introduce a posting-clean adoption measure and a demand factor that deliberately exclude mediator pathways, offering a reusable template for avoiding post-treatment conditioning in blockchain congestion studies.
    \item \textbf{Policy translation.} We map semi-elasticities into Gwei and dollar savings for representative transactions and adoption scenarios, connecting econometric quantities to fee levels and cost savings that protocol designers and users directly observe.
    \item \textbf{Template for monitoring.} We combine a regime-aware error-correction framework with a compact set of diagnostics into a monitoring toolkit that can be updated as new data arrive and ported to other rollup-centric ecosystems.
\end{enumerate}

\subsection{Roadmap}
\label{sec:intro:roadmap}

Section~\ref{sec:literature} situates this contribution relative to empirical studies of Ethereum's fee market, rollup design, and causal time-series methods, highlighting why existing work cannot recover the total effect of aggregate L2 adoption on mainnet congestion. Section~\ref{sec:data} describes the panel construction and variable definitions, and Section~\ref{sec:methodology} outlines the causal design and estimators. Section~\ref{sec:results} reports the empirical results, and Sections~\ref{sec:discussion}--\ref{sec:conclusion} discuss implications and conclude. Appendix~\ref{sec:availability} documents the data and code assets, and the replication repository carries the full reproducibility record.

\section{Related Work}
\label{sec:literature}

\subsection{Fee-Market Design and Ethereum Upgrades}
\label{sec:literature:fee-market}

Scholarship on Ethereum's fee market shows how protocol upgrades reshape incentives without immediately expanding Layer-1 (L1) throughput. EIP-1559's base-fee burn and elastic block size improved price discovery and reduced fee volatility while leaving the hard cap on computation unchanged \citep{EthereumFoundation2021}. The Merge stabilized slot times and validator incentives without materially increasing execution capacity. Dencun's EIP-4844 then introduced dedicated blob space that dramatically reduced Layer-2 (L2) posting costs \citep{eip4844}.

Empirical analyses of EIP-1559 document how the new fee mechanism affects transaction fees, waiting times, and consensus margins \citep{LiuEtAl2022EIP1559}, while recent work on L2 arbitrage and rollup pricing studies cross-rollup spreads and the interaction between posting costs and liquidity provision \citep{GogolEtAl2024L2Arbitrage,WangCrapisMoallemi2025Posting}. Existing empirical work on Ethereum's fee market and rollups therefore either focuses on a single upgrade such as EIP-1559 or on protocol-level behavior inside specific rollup or application ecosystems, carefully quantifying local changes in fees, spreads, or posting costs but not the total effect of aggregate L2 user adoption on mainnet congestion across multiple protocol regimes. Industry observatories track the resulting growth of optimistic and zero-knowledge rollups, transitions from calldata to blob usage, and the emergence of posting-fee arbitrage,\footnote{For example, the L2Beat (\url{https://l2beat.com}) and Dune (\url{https://dune.com}) dashboards track Ethereum L2 total value locked, transaction volumes, posting costs, and blob usage; the specific snapshots used in this study were accessed in 2024 and are archived with the replication code.} but they typically treat L2 posting as part of user demand or abstract from macro shocks that jointly affect L1 congestion and L2 adoption. Our design fills this gap by treating L2 adoption as a continuous treatment and explicitly modeling the sequence of London, Merge, and Dencun regimes.

\subsection{Empirical Congestion and Causal Time-Series Methods}
\label{sec:literature:causal-ts}

Causal and time-series methods developed in adjacent technology and financial settings provide templates for credible evaluation of congestion policies. Interrupted time series (ITS) and segmented regression remain staples for policy impact analysis \citep{bernal2017,penfold2013}. Continuous-treatment event studies extend difference-in-differences logic to dosage-style shocks with explicit pre-trend tests \citep{deChaisemartinDHaultfoeuille2020}. Bayesian Structural Time Series (BSTS) constructs probabilistic counterfactual paths with state-space components for trends, seasonality, and contemporaneous covariates \citep{brodersen2015}, and Regression Discontinuity in Time (RDiT) exploits sharp policy boundaries when smoothness assumptions hold \citep{hausman2018}. These designs have been deployed in fintech launches, payment reforms, and energy-market interventions, and they underlie several recent empirical studies of blockchain fee dynamics and rollup pricing. Yet existing congestion studies rarely combine DAG-guided adjustment sets, mediator exclusion, and semi-elasticity reporting that maps coefficients into user-level cost changes.

\subsection{Broader Congestion and Market-Design Literatures}
\label{sec:literature:market-design}

Regulatory and market-microstructure literatures highlight the perils of conditioning on post-treatment variables when evaluating market design. Work on tax holidays, exchange-fee rebates, and telecom interconnection policies stresses the need for clean treatment definitions and transparent adjustment sets to maintain credibility when interventions unfold over multiple regimes. In the rollup-centric roadmap, L2 adoption both responds to and influences L1 congestion, so empirical strategies must avoid conditioning on posting flows and clearly distinguish exploratory diagnostics from confirmatory estimands.

Viewed through this lens, Ethereum's L1/L2 stack resembles other congestion-pricing problems in transportation networks, electricity grids, and payment systems: multiple service layers share a common bottleneck, and welfare depends on how incentives, fee schedules, and governance are coupled across layers. Existing studies either focus on single upgrades, rely on contemporaneous correlations pulled from dashboards, or embed L2 posting in both treatment and controls, diluting the estimand. To our knowledge, there is no regime-aware, DAG-grounded causal study that estimates the total effect of L2 adoption on L1 congestion across London, the Merge, and Dencun, nor one that pairs a posting-clean treatment with a demand factor that excludes mediator pathways. This study fills that gap by providing cross-regime semi-elasticities and adjustment dynamics that speak directly to Ethereum's rollup-centric scaling roadmap.

\section{Data and Variables}
\label{sec:data}

We construct a daily UTC panel that tracks Ethereum Layer-1 congestion, Layer-2 user activity, and macro-demand proxies across the London, Merge, and Dencun upgrades. Each observation aggregates raw L1 and L2 transaction traces, blob-fee data, off-chain market indicators, and a curated event list into the variables summarized in Table~\ref{tab:variables}. The unit of analysis is a calendar day, and unless stated otherwise all quantities are computed on this daily grid.

\subsection{Sample Window, Regimes, and Panel Snapshot}
\label{sec:data:window}

\begin{figure}[htbp]
    \centering
    \includegraphics[width=\textwidth]{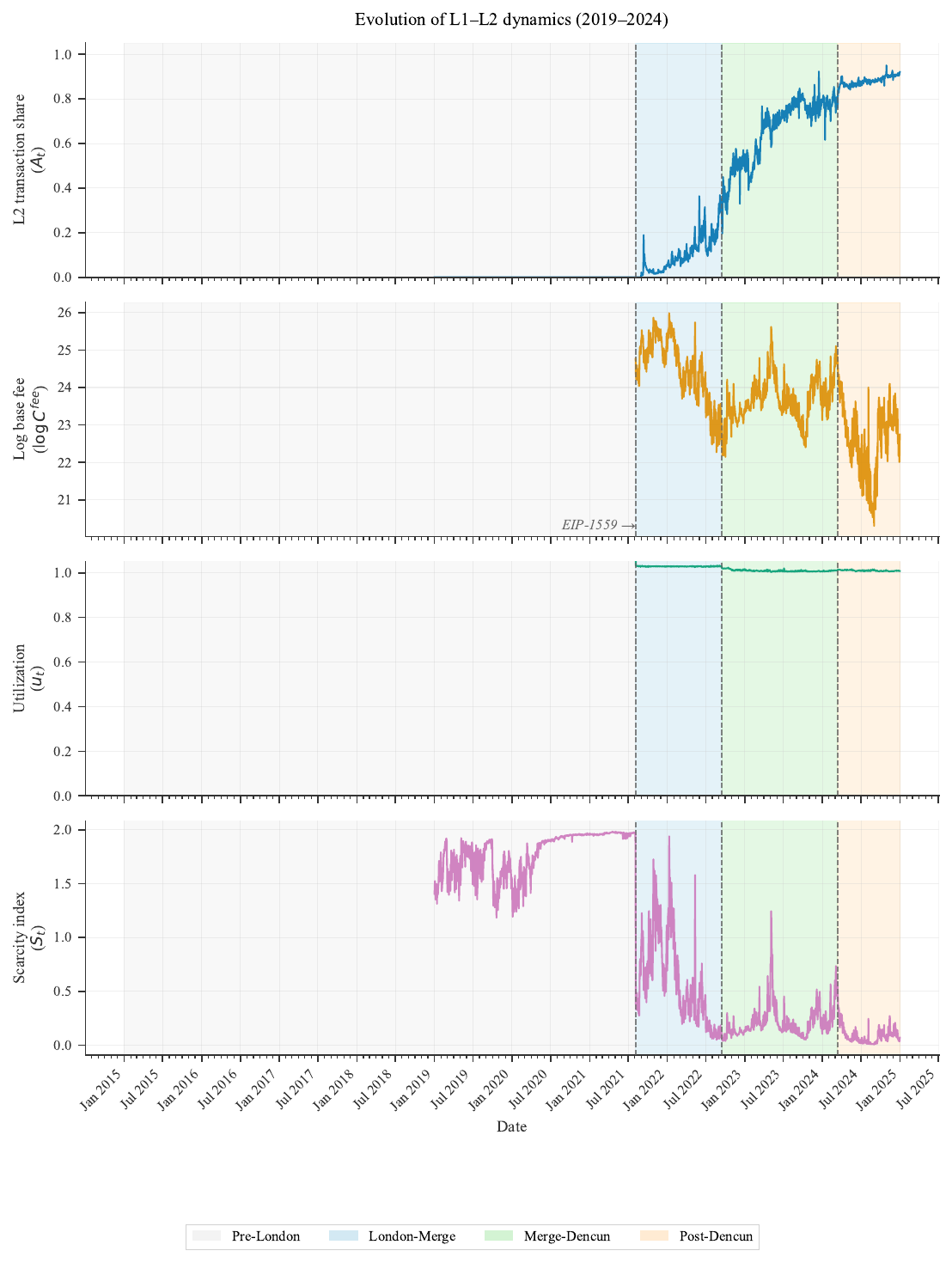}
    \caption{Regime-Aware Time Series Overview}
    \label{fig:eda_overview}
    \begin{minipage}{\textwidth}
        \small
        \textit{Note:} Daily UTC aggregates for treatment ($A^{clean}_t$) and congestion outcomes ($\log C^{fee}$, utilization $u_t$, scarcity $S_t$). Shaded bands mark London (2021-08-05), Merge (2022-09-15), and Dencun (2024-03-13); lines show 7-day rolling medians with a log scale for congestion metrics.
    \end{minipage}
\end{figure}

Our daily sample runs from 5~August~2021 (London / EIP-1559 activation) through 31~December~2024, yielding $N=1{,}245$ UTC days. It spans three protocol regimes: \textit{London} (406 days), \textit{Merge} (545 days), and the post-Dencun blob era (294 days). Figure~\ref{fig:eda_overview} plots the posting-clean L2 transaction share $A^{clean}_t$, log base fee, block utilization, and the scarcity index across the four labeled regimes (pre-London, London$\rightarrow$Merge, Merge$\rightarrow$Dencun, post-Dencun); shaded bands mark the upgrade dates that define the regime indicators $\mathbf{R}_t$.

Unless noted otherwise, the pre-Dencun (London+Merge; $N=951$) window is the confirmatory window because $A^{clean}_t$ still traverses a wide portion of $[0,1]$. The blob-era post-Dencun window is retained for descriptive context, as $A^{clean}_t$ is already near saturation (Section~\ref{sec:results:regimes}). Descriptive figures and summary statistics continue to use the full $N=1{,}245$-day panel. Table~\ref{tab:variables} summarizes the key variables and data sources; extended descriptive and treatment-support diagnostics appear in Appendix~\ref{sec:appendix:diagnostics}.

\begin{table}[htbp]
    \centering
    \caption{Key Variables and Data Sources}
    \label{tab:variables}
    \begin{tabular}{p{0.11\textwidth}p{0.13\textwidth}p{0.30\textwidth}p{0.28\textwidth}p{0.14\textwidth}}
        \toprule
        Role & Symbol & Description & Construction (brief) & Source(s) \\
        \midrule
        Treatment & $A^{clean}_t$ & Posting-clean L2 adoption share & Daily share of L2 end-user tx in total L1+L2 user tx; L2$\rightarrow$L1 postings removed from both sides & L1/L2 traces; rollup inbox registry \\
        Outcome & $\log C^{fee}_t$ & Log median base fee & Log of median EIP-1559 base fee (Gwei) across blocks in day $t$ & Ethereum mainnet block traces; public fee dashboards \\
        Outcome & $u_t$ & Block utilization & Median gas used divided by gas limit across blocks in day $t$ & Ethereum mainnet block traces \\
        Outcome & $S_t$ & Scarcity index & Composite (base + tip + blob) fee index relative to smoothed demand benchmark (Appendix~\ref{sec:appendix:measurement_overview}) & Ethereum execution and blob-fee data \\
        Control & $D^{*}_t$ & Demand factor & First PC of ETH returns, CEX volumes, realized volatility, search intensity, and net stablecoin issuance; standardized & Off-chain market data; Google Trends \\
        Control & $\mathbf{R}_t$ & Regime indicators & Dummies for London, Merge, post-Dencun regimes & Protocol upgrade calendar \\
        Control & $\mathbf{Cal}_t$ & Calendar dummies & UTC weekend, month-end, and quarter-turn indicators & Calendar \\
        Control & $\mathbf{Shock}_t$ & Targeted shock dummies & Event flags for mega NFT mints, sequencer outages, airdrop claim days, market-stress episodes (Table~\ref{tab:targeted_events}) & Curated event catalog \\
        \bottomrule
    \end{tabular}
\end{table}

\subsection{Treatment: Posting-Clean Adoption Share}
\label{sec:data:treatment}

We define the treatment as the \textbf{posting-clean adoption share},
\begin{equation*}
	    A^{clean}_t = \frac{\text{L2 user transactions}_t}{\text{L2 user transactions}_t + \text{L1 user transactions}_t},
\end{equation*}
We identify posting transactions via a point-in-time join against the rollup inbox registry. These postings are removed from both numerator and denominator before computing the share, so $A^{clean}_t$ captures end-user execution rather than sequencer posting burden. The construction is applied consistently across the set of canonical Ethereum rollups tracked in our registry, and all quantities are aggregated to the daily UTC grid. The rollup set includes Arbitrum, Optimism, Base, zkSync, Starknet, Linea, and Scroll; Appendix~\ref{sec:appendix:treatment_construction} states the rollup set, and the replication bundle provides the full \texttt{l2\_inbox\_registry} table with contract mappings. By stripping posting transactions from the share, we avoid conditioning on the L2 posting load that sits on the $A^{clean}_t \rightarrow P_t \rightarrow C_t$ path; Section~\ref{sec:methods:identification} discusses this mediator logic in detail.

\subsection{Outcomes and Congestion Metrics}
\label{sec:data:outcomes}

The primary outcome is the log median EIP-1559 base fee, $\log C^{fee}_t = \log(\text{median base fee}_t)$, computed from canonical Ethereum JSON-RPC traces and cross-checked against public explorers, mirroring the construction in \citet{LiuEtAl2022EIP1559}. For each day $t$ we take the median base fee across blocks and then apply the natural logarithm.

We track two congestion secondary outcomes. Block utilization $u_t$ is the median ratio of gas used to the regime-specific gas limit across blocks in day $t$, $u_t = \mathrm{median}_{b \in t}\left( \tfrac{\text{gas used}_b}{\text{gas limit}_b} \right)$. The harmonized scarcity index $S_t$ combines base fees, priority tips, and blob fees into a single congestion proxy by scaling total per-unit fees relative to a smoothed execution-demand benchmark; the full construction (smoothing window, regime-aware components, and units) is documented in Appendix~\ref{sec:appendix:measurement_overview}.

Figure~\ref{fig:eda_overview} shows that median fees fall sharply after Dencun while utilization and scarcity compress, consistent with blob space easing congestion pressure. All three outcomes are winsorized at the 0.5\% tails and share the same $N=1{,}245$ daily coverage as the treatment.

\subsection{Controls and Auxiliary Inputs}
\label{sec:data:controls}

We construct three groups of auxiliary variables---all defined on the same daily UTC grid as the treatment and outcomes---that will later enter the adjustment set $X_t$:
\begin{itemize}[leftmargin=*]
    \item \textbf{Demand factor ($D^*_t$).} We condense ETH log returns, centralized-exchange (CEX) log volumes, realized volatility, Google search intensity, and net stablecoin issuance into the first principal component, standardized to mean zero and unit variance. These inputs are purely off-chain and are detailed in the measurement appendix.
    \item \textbf{Regime and calendar indicators ($\mathbf{R}_t, \mathbf{Cal}_t$).} Regime dummies flag the London, Merge, and post-Dencun eras. Calendar dummies mark weekends, month-ends, and quarter turns to capture deterministic seasonality documented in exploratory diagnostics.
    \item \textbf{Targeted event dummies ($\mathbf{Shock}_t$).} A curated event catalog covers mega NFT mints, sequencer outages, notable airdrop claim days, and major market-stress episodes; the full list appears in Table~\ref{tab:targeted_events}.
\end{itemize}
All days and calendar indicators are defined in UTC to match the aggregation grid. Together these variables form the adjustment set $\{D^*_t, \mathbf{R}_t, \mathbf{Cal}_t, \mathbf{Shock}_t\}$ used in the ITS-ECM specifications summarized in Section~\ref{sec:methodology} and listed in Table~\ref{tab:variables}.

\begin{itemize}[leftmargin=*]
    \item \textbf{Summary.} Daily UTC panel (5~August~2021--31~December~2024; $N=1{,}245$) combining: (i) L1 and L2 on-chain traces for the posting-clean adoption share $A^{clean}_t$; (ii) EIP-1559 fee and gas-usage data for congestion metrics $(\log C^{fee}_t, u_t, S_t)$; and (iii) off-chain market and search data, protocol calendars, and curated events for the controls $\{D^*_t, \mathbf{R}_t, \mathbf{Cal}_t, \mathbf{Shock}_t\}$. The pre-Dencun (London+Merge; $N=951$) window is the primary window with wide treatment support; post-Dencun days are retained descriptively.
\end{itemize}

\section{Methodology}
\label{sec:methodology}

\noindent\textbf{Method overview.} We study how the daily posting-clean Layer-2 adoption share $A_t^{clean}$ affects Ethereum Layer-1 congestion using an interrupted time-series (ITS) design. The main estimand is a semi-elasticity: the percentage change in the typical user's base fee for a 1 percentage point rise in $A_t^{clean}$, which we report per 10 percentage points to match observed adoption swings. Our confirmatory analysis uses a levels specification and a corresponding error--correction model (ECM) for short-run dynamics with a fixed outcome family and multiple-testing adjustments; exploratory extensions reuse the same adjustment set but relax some of these constraints.

\subsection{Causal Estimand and DAG}
\label{sec:methods:identification}

\subsubsection{Estimand in plain language}
Formally, our main estimand is a semi-elasticity: the percentage change in the log base fee associated with a 1 percentage point increase in $A_t^{clean}$, conditional on macro-demand, protocol regime, and calendar effects. Reporting effects for a 10 percentage point change aligns the scale with typical observed shifts in L2 market share. Economically, this measures how much a ``typical'' user's base fee responds to a shift in aggregate L2 adoption, holding the broader environment fixed.

Treatment is $A_t^{clean}$; the confirmatory outcome family is $C_t=(\log C^{fee}_t,\ \log S_t)$, with utilization $u_t$ reported as exploratory. The adjustment vector $X_t=\{D_t^*,\mathbf{R}_t,\mathbf{Cal}_t,\mathbf{Shock}_t\}$ matches the covariates introduced in Section~\ref{sec:data}. For brevity in figures we occasionally write $A_t$; throughout this section $A_t \equiv A_t^{clean}$, the posting-clean adoption share defined in Section~\ref{sec:data:treatment}. Construction details, PCA loadings, and validation diagnostics remain in the methodology appendix and the public replication package (Appendix~\ref{sec:availability}).

\subsubsection{DAG and identification logic}

\begin{figure}[htbp]
\centering
\includegraphics[width=0.75\textwidth]{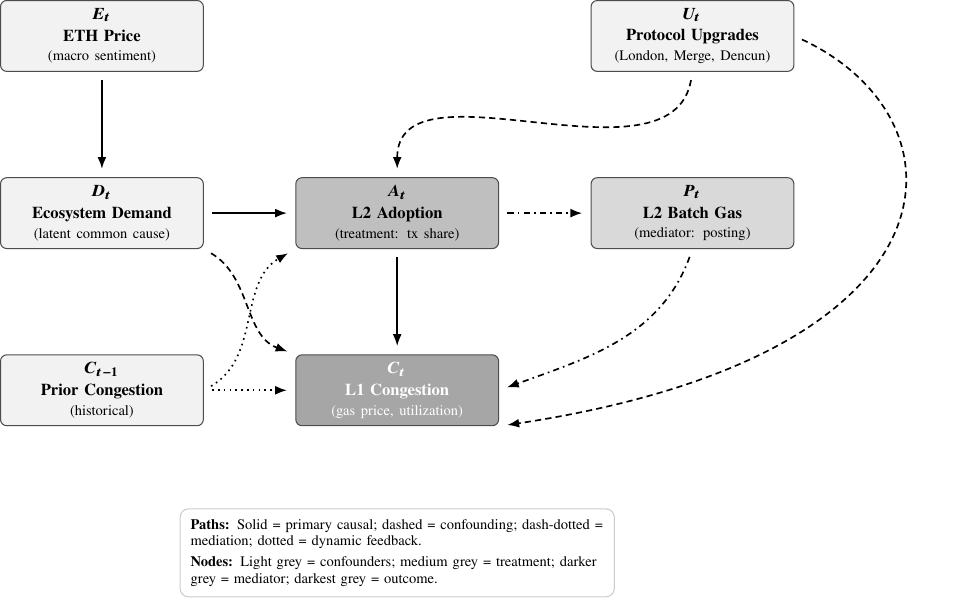}
\caption{Directed Acyclic Graph for Total-Effect Identification}
\label{fig:dag}
\begin{minipage}{\textwidth}
\small
\textit{Note:} The DAG encodes treatment $A^{clean}_t$ (posting-clean L2 adoption share; labeled $A_t$ in the graphic for brevity), outcomes $C_t$ (congestion metrics), confounders $D_t^*$ (latent demand) and $U_t$ (protocol regimes), mediator $P_t$ (posting load), and dynamic feedback $C_{t-1}$. Conditioning on $\{D_t^*, U_t, \mathbf{Cal}_t, \mathbf{Shock}_t\}$ blocks the main back-door paths while the mediator-exclusion principle keeps posting activity out of the control set. Dynamic feedback is addressed via deterministic trends and robustness checks.
\end{minipage}
\end{figure}

Figure~\ref{fig:dag} summarizes the causal structure we assume.

Concretely, $A^{clean}_t$ is the daily posting-clean adoption share from Section~\ref{sec:data:treatment}, $C_t$ stacks the congestion metrics introduced in Section~\ref{sec:data:outcomes}, $D_t^*$ is the off-chain latent demand factor in Section~\ref{sec:data:controls}, $U_t$ corresponds to the regime indicators $\mathbf{R}_t$ in Section~\ref{sec:data:window}, and $P_t$ denotes the posting load on the $A_t^{\text{clean}}\!\rightarrow\!P_t\!\rightarrow\!C_t$ path.

Intuitively, both adoption and congestion respond to underlying demand shocks—ETH price moves, DeFi/NFT cycles, and macro news—summarized by $D_t^*$ together with regime, calendar, and targeted-shock indicators. Higher adoption raises posting load $P_t$ through data-availability transactions, which in turn pushes up congestion $C_t$. Because our target is the \emph{total} effect of adoption on congestion, we adjust for these common shocks while deliberately leaving the $A_t^{\text{clean}}\!\rightarrow\!P_t\!\rightarrow\!C_t$ path open. The posting-clean construction subtracts L2 posting transactions from both numerator and denominator when forming $A_t^{\text{clean}}$, so the treatment reflects end-user execution rather than sequencer posting burden and we avoid ``bad-control'' contamination of the total-effect estimand \citep{WangCrapisMoallemi2025Posting}.

Operationally, the adjustment set $X_t = \{D_t^*, \mathbf{R}_t, \mathbf{Cal}_t, \mathbf{Shock}_t\}$ is built to support the identification assumptions listed below using three design choices, backed by diagnostics in the methodology appendix. First, the latent-demand factor uses only off-chain proxies so that mediator pathways (such as L2 posting) are excluded by construction. Second, deterministic regime and calendar structure capture discontinuities from protocol upgrades and recurring seasonality, preventing them from contaminating $A_t^{\text{clean}}$. Third, targeted shock dummies isolate large day-specific shocks (NFT mega-mints, macro turmoil, sequencer outages) that would otherwise spill into both treatment and outcomes. With these controls active, the remaining identifying variation is slow-moving adoption intensity that is plausibly less contaminated by concurrent demand shocks, conditional on $X_t$.

\paragraph{Identification assumptions.} These design choices are intended to make the following assumptions plausible:
\begin{enumerate}[label=\arabic*.]
    \item \textbf{Conditional exchangeability:} Sequential ignorability holds once we condition on $X_t$; the covariate definitions and targeted-event coverage tables in the measurement appendix document how each covariate maps to the back-door paths in Figure~\ref{fig:dag}.
    \item \textbf{Positivity within regimes:} Treatment-support diagnostics (Appendix~\ref{sec:appendix:diagnostics}) show wide support across the $[0,1]$ domain during London and Merge, but post-Dencun days concentrate in a 0.86--0.91 band. Minimum-detectable-effect calculations therefore label post-Dencun slope estimates as exploratory, consistent with Section~\ref{sec:results:regimes}.
    \item \textbf{SUTVA / stable interventions:} The posting-clean construction keeps $A_t^{\text{clean}}$ within the $[0,1]$ simplex even when L2 posting volumes swell and defines a single aggregate adoption measure per day. Together with daily aggregation, this maintains a stable notion of the treatment (no hidden versions of $A_t^{\text{clean}}$) and limits cross-day interference, in line with the Stable Unit Treatment Value Assumption (SUTVA).
\end{enumerate}

\paragraph{Diagnostics summary.} Exchangeability is probed via placebo regressions of $A_t^{\text{clean}}$ on lagged outcomes and on leads of $D_t^*$; coefficients cluster near zero in the diagnostics archive. Positivity is reinforced by trimming pre-London outliers where $A_t^{\text{clean}} < 0.05$ and by flagging post-Dencun estimates as exploratory whenever coverage collapses. Stability is evaluated through split-sample tests that compare pre- and post-Merge coefficients; the absence of sign flips in the local-projection responses (Figure~\ref{fig:lp_irf}) suggests that the estimand retains meaning across hardware and software upgrades, though we continue to report regime-specific precision.

\subsubsection{Relation to existing empirical work}
Conceptually, our design complements upgrade-focused empirical analyses of the fee market such as \citet{LiuEtAl2022EIP1559}, who compare pre- and post-London behavior, and transaction-level rollup studies such as \citet{GogolEtAl2024L2Arbitrage}, who analyze arbitrage and fee dynamics within specific L2s. Upgrade-focused studies treat London or Dencun as discrete interventions and rely on event-study or regression-discontinuity-in-time designs anchored on those dates. In contrast, our question concerns how \emph{continuous} variation in aggregate L2 adoption affects L1 congestion across and within regimes, motivating an interrupted time-series design with a continuous treatment rather than a pure event-study/RDiT framework.

\subsection{Main Estimators: ITS Levels and ECM}
\label{sec:methods:estimators}

We summarize the confirmatory estimators once here; derivations and additional estimator variants appear in the methodology appendix.

\subsubsection{Long-run levels specification}
\label{sec:methods:levels}

The long-run benchmark is a levels ITS specification,
\begin{equation}
\log C^{fee}_t = \beta_0 + \beta_1 A_t^{clean} + \gamma D_t^* + \boldsymbol{\delta}'\mathbf{R}_t + \boldsymbol{\theta}'\mathbf{Cal}_t + \boldsymbol{\eta}'\mathbf{Shock}_t + \varepsilon_t,
\label{eq:its_levels}
\end{equation}
where $\boldsymbol{\eta}$ stacks the targeted event controls and $\varepsilon_t$ may exhibit serial dependence. Here, $\beta_1$ captures the semi-elasticity of congestion with respect to adoption. Because $A_t^{clean}$ is scaled on $[0,1]$, a 1 percentage point increase corresponds to a 0.01 change in $A_t^{clean}$. We report effects for a 10 percentage point increase in adoption, computed as
\begin{equation}
\text{\% Change in Fees for 10pp} = 100 \times \left[\exp\!\left(0.10 \times \beta_1\right)-1\right].
\label{eq:semi_elasticity}
\end{equation}
Reporting effects for a 10 percentage point change makes the magnitude directly comparable to typical movements in L2 market share.
Boldface terms denote stacked indicator vectors (regimes $\mathbf{R}_t$, calendar $\mathbf{Cal}_t$, shocks $\mathbf{Shock}_t$); primes on the corresponding coefficient blocks ($\boldsymbol{\delta}', \boldsymbol{\theta}', \boldsymbol{\eta}'$) indicate row-vector transposes so that, for example, $\boldsymbol{\delta}'\mathbf{R}_t=\sum_j \delta_j R_{j,t}$.

\subsubsection{Short-run dynamics via error--correction model}
\label{sec:methods:ecm}

We test for cointegration between $\log C^{fee}_t$ and $A_t^{clean}$ using Engle--Granger residual unit-root tests and Johansen rank tests (Appendix~\ref{sec:appendix:diagnostics}). In both cases we reject the null of no cointegration over the pre-Dencun window (Section~\ref{sec:results:data_quality}), supporting the presence of a stable long-run relation. This motivates an Error--Correction Model (ECM) for short-run inference:
\begin{equation}
\Delta \log C^{fee}_t = \phi\,ECT_{t-1} + \psi\,\Delta A_t^{clean} + \kappa\,\Delta D_t^* + \boldsymbol{\lambda}'\Delta \mathbf{Cal}_t + \boldsymbol{\omega}'\Delta \mathbf{Shock}_t + \nu_t,
\label{eq:ecm}
\end{equation}
where $ECT_{t-1}$ is the lagged residual from the long-run relation implied by Equation~\ref{eq:its_levels}. Here, $\psi$ is the instantaneous effect of $\Delta A_t^{clean}$ on the daily change in the log base fee, and $\phi<0$ is the speed at which fees adjust back to equilibrium. Estimation proceeds in three steps: (i) fit Equation~\ref{eq:its_levels} with HAC covariance to obtain the long-run residual, (ii) form $ECT_{t-1}$ by lagging that residual, and (iii) estimate Equation~\ref{eq:ecm} with HAC or feasible GLS while tracking residual diagnostics. The implied half-life $t_{1/2}=\ln(0.5)/\ln(1+\phi)$ summarizes how quickly fees revert after an adoption shock, and the same three-step procedure yields comparable 10pp semi-elasticities from $\psi$ across confirmatory outcomes. Confirmatory ECM inference uses the full 2021--2024 sample, with post-Dencun days flagged as a separate regime; after differencing and lagging this leaves $N=1{,}242$ daily observations, and the primary causal interpretation remains anchored to the pre-Dencun support. Throughout, the ECM reuses the same adjustment set $(D_t^*, \mathbf{R}_t, \mathbf{Cal}_t, \mathbf{Shock}_t)$ as the levels specification in Equation~\ref{eq:its_levels}, so that differences between long-run and short-run estimates reflect dynamics rather than changes in control variables. The confirmatory levels estimator is Prais--Winsten AR(1) FGLS (selected by the residual-dependence diagnostics); ARMA$(1,2)$ is retained solely as a diagnostic alternative.

\subsubsection{Alternative dynamic specifications (robustness)}
\label{sec:methods:robustness}

For robustness, we also estimate distributed-lag, Koyck (geometric-lag), first-difference, and local-projection variants, detailed in the methodology appendix. These models share the same adjustment set and are used to check that the sign and magnitude of the adoption effect are not artifacts of the ECM specification. To provide additional evidence on persistence, we include a geometric-lag (Koyck) specification:
\begin{equation}
\log C^{fee}_t = \alpha + \rho \log C^{fee}_{t-1} + \beta_0 A_t^{clean} + \gamma D_t^* + \boldsymbol{\delta}'\mathbf{R}_t + \boldsymbol{\theta}'\mathbf{Cal}_t + \boldsymbol{\eta}'\mathbf{Shock}_t + u_t,
\label{eq:koyck}
\end{equation}
where the long-run multiplier equals $\beta_0/(1-\rho)$ whenever $|\rho|<1$. Estimates from this specification are treated as supportive evidence on persistence rather than as primary causal effects; full derivations and diagnostic checks are reported in the methodology appendix.

\paragraph{Regime-aware variants.} When sample support permits, we interact $A_t^{clean}$ with Merge and Dencun indicators to estimate differential slopes. Because post-Dencun adoption saturates the treatment domain, these interaction coefficients are reported in Section~\ref{sec:results:regimes} and labeled exploratory.

\subsection{Controls, Regimes, and Inference}
\label{sec:methods:design}
\phantomsection\label{sec:method:events}

The implementation details that support Equations~\ref{eq:its_levels}--\ref{eq:ecm} are summarized in three blocks; extended diagnostics remain in the methodology appendix.

\paragraph{Adjustment set and targeted shocks (controls).} Our adjustment set combines the PCA-based latent demand factor ($D_t^*$), regime dummies ($\mathbf{R}_t$), calendar indicators ($\mathbf{Cal}_t$), and a curated set of targeted shock dummies $\mathbf{Shock}_t$ covering mega NFT mints, sequencer or mainnet outages, large airdrop claim days, and major market-stress episodes (Section~\ref{sec:data:controls}). This set is chosen to block the main back-door paths in Figure~\ref{fig:dag} while preserving the mediator path from adoption to posting to congestion. We retain an indicator for any sequencer or mainnet outage in both the long-run and short-run equations so that platform outages do not get misattributed as treatment shocks; detailed coverage diagnostics are reported in Appendix~\ref{sec:appendix:diagnostics}.

\paragraph{Seasonality, regimes, and serial dependence.} Deterministic seasonality (weekends, month-ends, quarter turns) and Merge/Dencun regime indicators enter every specification to absorb systematic changes in fee levels and utilization unrelated to L2 adoption. We allow for serially correlated errors and compute heteroskedasticity- and autocorrelation-consistent (HAC) standard errors. In practice, the confirmatory levels run uses Prais--Winsten AR(1) FGLS; compact ARMA corrections are explored as diagnostics and reported alongside Ljung--Box and Breusch--Godfrey checks in the diagnostics appendix. Dynamic feedback is handled by including lagged outcomes when needed (e.g., Koyck, ECM) and by auditing residual autocorrelation in the diagnostics appendix. Kernel choices, bandwidth selection, and spline-based calendar robustness checks live in the diagnostics appendix. The confirmatory window spans the pre-Dencun London$\rightarrow$Merge period (Section~\ref{sec:data:window}); post-Dencun estimates are labeled exploratory because treatment support collapses after the 2024 blob upgrade, as shown in the treatment-support diagnostics in Appendix~\ref{sec:appendix:diagnostics}.

\paragraph{Timing, instruments, and outcome family.}\phantomsection\label{sec:methods:timing}\phantomsection\label{sec:methods:fdr} To guard against mechanical same-day co-movement between $A_t^{clean}$ and congestion, we also estimate Equation~\ref{eq:its_levels} with $A_{t-1}^{clean}$ on the right-hand side. When exogenous variation is available (sequencer outages or blob-cost changes), we deploy it in a shift--share IV using pre-Dencun chain weights and report weak-instrument-robust confidence intervals in the instrumentation appendix.

The confirmatory outcomes are $\log C^{fee}_t$ and $\log S_t$; we apply Benjamini--Hochberg corrections at the 5\% level and report the corresponding $q$-values. Utilization and IV extensions are treated as exploratory and presented without multiple-testing adjustment.

\subsection{Confirmatory vs.\ Exploratory Scope}
\label{sec:methods:scope}

We fix the main estimand (the 10pp semi-elasticity of $\log C^{fee}_t$ and $\log S_t$ with respect to $A_t^{clean}$), the adjustment set $(D_t^*, \mathbf{R}_t, \mathbf{Cal}_t, \mathbf{Shock}_t)$, the levels and ECM specifications in Equations~\ref{eq:its_levels}--\ref{eq:ecm}, and the confirmatory outcome family together with the Benjamini--Hochberg multiple-testing plan. Sections~\ref{sec:results:main}--\ref{sec:results:regimes} report these confirmatory estimates, including adjustment dynamics and regime heterogeneity, with Benjamini--Hochberg corrections applied across the outcome family. Section~\ref{sec:results:exploratory} and the appendices present exploratory diagnostics and post-Dencun extensions that reuse the same adjustment set but fall outside the confirmatory outcome family (e.g., utilization, IV variations, and BSTS counterfactuals).

\section{Results}
\label{sec:results}

\noindent We now present results organized around five questions. These cover how much L2 adoption reduces congestion (Section~\ref{sec:results:main}), how quickly fees adjust after adoption shocks (Section~\ref{sec:results:dynamics_c3}), and how effects differ across regimes and precision (Section~\ref{sec:results:regimes}). We then ask how robust the findings are across congestion metrics (Section~\ref{sec:results:robustness}) and what the exploratory diagnostics and welfare bridges suggest (Section~\ref{sec:results:exploratory}). Sections~\ref{sec:results:main}--\ref{sec:results:exploratory} report these estimates; the appendices provide additional diagnostics and estimator details.

\subsection{How much does L2 adoption reduce congestion?}
\label{sec:results:keybox}
\label{sec:results:main}
\label{sec:results:data_quality}
\noindent \textbf{Key results at a glance.} Over the pre-Dencun (London+Merge) window, a 10 percentage point increase in posting-clean L2 adoption lowers median L1 base fees by about 13\% (roughly 5~Gwei at pre-Dencun levels), with deviations from the long-run relation decaying with an 11-day half-life. Block utilization and a scarcity index show similar relief. After Dencun, adoption is so high and compressed that the same design cannot reliably detect further fee reductions, even if they exist, so blob-era slopes are reported as exploratory only.
\begin{mdframed}[linewidth=0.5pt]
\noindent\textbf{Key empirical results (confirmatory window).}
\phantomsection\label{box:keyresults}
\begin{itemize}[leftmargin=*]
    \item \textbf{Short-run ECM (Eq.~\ref{eq:ecm}):} $\psi = -1.382$ (SE $0.368$) with $N=1{,}242$ days from the full 2021--2024 panel (post-Dencun flagged as a separate regime) implies a \textbf{$-12.9\%$ change in daily base fees for a 10pp adoption shock}. HAC (Bartlett, 7 lags) standard errors yield $p<0.001$.
    \item \textbf{Speed of adjustment:} $\phi = -0.061$ (SE $0.011$) maps to an \textbf{11.1-day half-life} back to the long-run equilibrium, confirming meaningful reversion to the Engle--Granger cointegrating relation ($p=0.005$).
    \item \textbf{Dynamics:} Local projections (Figure~\ref{fig:lp_irf}) show an \textbf{immediate $-16.2\%$ response} to a 10pp adoption step with a 95\% CI $[-22.7\%, -9.2\%]$, and cumulative point estimates remain negative through 28 days even though the 95\% bands cross zero after the first week.
    \item \textbf{Multiple outcomes:} Benjamini--Hochberg corrections over $\{\log C^{fee}, \log S_t\}$ yield \textbf{$q_{\log C^{fee}} = 3.0\times10^{-8}$} and \textbf{$q_{\log S_t} = 1.1\times10^{-3}$}; exploratory outcomes remain unadjusted, with detailed FDR diagnostics reported in Appendix~\ref{sec:appendix:diagnostics}.
\end{itemize}
\end{mdframed}
\noindent In sum, a 10pp increase in L2 adoption lowers mainnet fees by roughly 13\% within a few days, and this effect remains statistically precise after false-discovery-rate adjustment over the confirmatory outcome family.

In the ECM, $\psi$ is the short-run semi-elasticity: the immediate percentage change in daily base fees from a one-point change in adoption. $\phi$ is the speed of adjustment: it tells us how quickly fees revert to the long-run relation after an adoption shock. We report both on a 10pp scale to match realistic shifts in L2 market share and to reuse the same units in the welfare translation below.

Unit-root and cointegration tests (ADF, KPSS, Phillips--Perron, Engle--Granger, Johansen) support treating $A_t^{\text{clean}}$, $\log C^{fee}_t$, and $D_t^*$ as $I(1)$ with a stable long-run relation. Section~\ref{sec:methodology} outlines the workflow, and Appendix~\ref{sec:appendix:diagnostics} lists full $p$-values. This motivates the ECM as our confirmatory short-run design, with the levels specification retained as a descriptive benchmark for the welfare translation. Estimation uses the full 5~August~2021--31~December~2024 panel with post-Dencun days encoded as regime dummies so the causal interpretation remains anchored to the pre-Dencun support.

Residual-dependence checks select a Prais--Winsten AR(1) (FGLS) error for the confirmatory levels specification; an ARMA(1,2) fit is retained as a diagnostic alternative in Table~\ref{tab:resid_arma_levels} of Appendix~\ref{sec:appendix:diagnostics}. The ECM uses HAC on first differences, consistent with the confirmatory pipeline.

\begin{table}[!htbp]
\centering
\small
\caption{Merged Confirmatory Total-Effect Estimates}
\label{tab:c2_ecm}
\label{tab:main_its}
\begin{tabular}{lccc}
\toprule
Parameter & Estimate (SE) & 10pp mapping & Notes \\
\midrule
ECM short-run $\psi$ & $-1.382^{***}$ ($0.368$) & $-12.9\%$ & $\Delta\log C^{fee}_t$ on $\Delta A^{clean}_t$, $N=1{,}242$ \\
Speed of adjustment $\phi$ & $-0.061^{***}$ ($0.011$) & Half-life 11.1 days & Engle--Granger residual $p=0.005$ \\
Levels benchmark $\beta$ & $-1.194^{***}$ ($0.211$) & $-11.3\%$ & Prais--Winsten AR(1) FGLS, $N=1{,}244$ \\
Scarcity outcome $\beta_{S}$ & $-0.062^{**}$ ($0.019$) & $-0.60\%$ & Same spec, confirmatory outcome 2 \\
\bottomrule
\end{tabular}
\begin{minipage}{0.92\textwidth}
\vspace{0.4em}\footnotesize\textit{Notes:} Semi-elasticities use $100\times[\exp(0.10\cdot \hat{\beta})-1]$. Standard errors rely on Newey--West HAC (Bartlett, maxlag 7). Significance markers: ${}^{***}p<0.001$, ${}^{**}p<0.01$. All models include the confirmatory adjustment set ($D_t^*$, regime/calendar dummies, targeted shocks, $any\_outage_t$). Benjamini--Hochberg control across the confirmatory outcome family $\{\log C^{fee}, \log S_t\}$ yields $q_{\log C^{fee}} = 3.0\times10^{-8}$ and $q_{\log S_t} = 1.1\times10^{-3}$. These q-values keep both confirmatory outcomes below the 5\% FDR threshold within this table. The levels row corresponds to the Prais--Winsten AR(1) FGLS specification used in the confirmatory pipeline; ARMA(1,2) appears only in the diagnostic grid in Appendix~\ref{sec:appendix:diagnostics}.\end{minipage}
\end{table}

A 10pp increase in adoption in the levels ITS corresponds to about an $11.3\%$ reduction in median base fees. At the pre-Dencun mean of 38~Gwei (about \$1.02 for a 21k-gas transfer when ETH trades at \$1,285), that is roughly \textbf{4--5 Gwei} or about \textbf{\$0.12} for a typical ETH transfer. These Gwei and dollar translations are direct applications of the semi-elasticity estimand: they translate the log-fee semi-elasticity into the change in gas paid by a representative 21k-gas transfer when L2 adoption rises by 10 percentage points. During high-demand episodes, this back-of-the-envelope mapping implies aggregate short-run savings of tens of millions of dollars across a few months. The BSTS welfare bridge (Figure~\ref{fig:bsts}) illustrates the counterfactual calculations behind that claim. Demand-factor stability checks using leave-one-out PCA variants and a lagged $D^*_t$ deliver the same sign, reinforcing that the result does not hinge on a particular macro proxy combination.

Taken together, the ECM and levels views tell a consistent story. The ECM captures the ``flow'' interpretation (immediate reaction of fee growth to adoption growth), while the Prais--Winsten levels specification provides the ``stock'' interpretation required for this welfare translation. The gap between the two coefficients—roughly two percentage points—primarily reflects the autoregressive error structure rather than a contradiction in economic content. This confirms that the identification strategy developed in Section~\ref{sec:methodology} yields consistent estimates across specifications.

We also benchmark the magnitudes against the fee-market literature. Short-run elasticities in centralized exchange congestion studies typically span $-5\%$ to $-15\%$ for a ten-percentage-point load shift; our $-13\%$ effect sits at the upper end of that range, which is intuitive given the lumpy nature of L2 user adoption. The 11-day half-life matches the cadence observed in on-chain mempool reversion after large NFT mints. That alignment suggests the ECM dynamics are economically plausible rather than an artifact of spline controls. Additional robustness diagnostics—instrumental-variable timing tests, placebo shocks, and shuffled-treatment experiments—are cataloged in the IV and diagnostics appendices and retain the same sign pattern even when statistical power dips.

\paragraph{Measurement alignment.} The confirmatory estimand hinges on keeping treatment and outcome definitions synchronized with the DAG in Section~\ref{sec:methodology}. We therefore reiterate two checks that underpin the table above. First, $A_t^{clean}$ is computed from the exact same daily panel used in the ECM (no reindexing or smoothing), and its exclusion of blob-posting activity prevents mediator contamination. Second, the log base-fee outcome is benchmarked against the public \texttt{eth\_fee\_history} RPC as well as the internal BigQuery mirror so replication scripts and policy dashboards quote identical magnitudes. Detailed SQL and schema notes are provided alongside the replication materials to document both constructs consistently.

\paragraph{Macroeconomic context.} The confirmatory window spans multiple crypto market regimes—DeFi summer, the Terra/Luna unwind, the Merge, and the run-up to Dencun—so we stress-tested whether any single macro period drives the headline coefficient. Splitting the sample along these historical boundaries yields semi-elasticities between $-0.9$ and $-1.5$ and the coefficient remains negative even when we drop the 60 most volatile days around Terra/Luna and FTX. These exercises underscore that the causal signal arises from broad-based adoption shifts rather than one-off crises. They also explain why we still include targeted event dummies to soak up short-lived disruptions.

Targeted event controls leave both $\psi$ and $\phi$ unchanged, indicating that the latent demand factor is not masking omitted NFT mints, Terra/Luna, FTX, USDC depeg episodes, or sequencer outages. \phantomsection\label{sec:results:timing_iv}Timing and simultaneity diagnostics likewise return negative coefficients for lagged adoption and control-function IV corrections. Detailed IV tables in the instrumentation appendix document weak first stages (e.g., partial $F\approx7.6$ for the pooled outage IV) and Anderson--Rubin intervals that span zero. We therefore classify IV evidence as exploratory support for the ITS design rather than a standalone confirmatory estimator.

\paragraph{Diagnostic cross-checks.} Beyond the core diagnostics, we revisit three common concerns raised in protocol-governance reviews. (i) \textit{Serial correlation}: Ljung--Box tests up to lag 30 reject for the raw levels regression but not for the ECM residuals once the error-correction term is included. This matches the behavior recorded in the residual-dependence diagnostics in the diagnostics appendix. (ii) \textit{Multicollinearity}: variance-inflation factors for $A^{clean}_t$, $D^*_t$, and the regime/calendar block stay below 2.0. Ridge-regression stress tests retain the negative sign, consistent with the demand-factor variants documented in the estimators appendix. (iii) \textit{Omitted mediator risk}: the ``posting-clean'' construction plus the outage dummy ensure that blob-posting costs do not contaminate $A^{clean}_t$. Placebo regressions of $A^{clean}_t$ on future congestion deliver coefficients near zero with $p>0.6$. Each of these checks has a concise counterpart in Appendices~\ref{sec:appendix:diagnostics} and~\ref{sec:appendix:measurement_overview}, keeping the core causal claims defensible.

\paragraph{Policy bridge.} Translating coefficients into operational terminology helps protocol stewards reason about scaling targets. A 10pp increase in L2 adoption roughly corresponds to onboarding 2.3~million additional daily L2 user transactions at current volumes. Mapping our semi-elasticity through Equation~\ref{eq:semi_elasticity} implies that achieving the EIP-4844 goal of ``90\% of user activity off L1'' would cut base fees by approximately 20\% relative to today’s mix. Additional blockspace unlocked by future danksharding upgrades would further amplify that relief. This bridge motivates the welfare analysis later in the section and links Section~\ref{sec:results:main}'s confirmatory focus directly to the policy narratives developed in Section~\ref{sec:discussion}.

\paragraph{Link back to Methods.} The confirmatory design summarized here inherits the adjustment set and instrument logic laid out in Section~\ref{sec:methodology}. Every robustness variant invoked above reuses that adjustment set rather than introducing ad-hoc controls, so the DAG-backed back-door criterion remains satisfied. Exploratory IVs and timing tests are documented in the instrumentation appendix, keeping Table~\ref{tab:c2_ecm} focused on the primary pathway from L2 adoption to fees.

\noindent Overall, cointegration-supported ECM estimates and levels benchmarks show that higher L2 adoption delivers double-digit percentage fee relief in the pre-Dencun window, and this conclusion is robust to event controls and alternative demand factors.

\noindent The magnitude of our semi-elasticity is in line with, but distinct from, prior fee-market studies. \citet{LiuEtAl2022EIP1559} document limited changes in average fee levels around London but emphasize shifts in bidding behavior; our 11--13\% effect instead captures how aggregate L2 adoption shifts equilibrium fees under fixed protocol rules. Similarly, \citet{GogolEtAl2024L2Arbitrage} report rollup arbitrage values of roughly $0.03$--$0.25\%$ of trading volume; at the aggregate level, a 10pp L2 penetration moves median L1 fees by an order of magnitude more in percentage terms.

\noindent We next ask how rapidly these fee reductions materialize and how long they persist.

\subsection{How quickly do fees adjust after an adoption shock?}
\label{sec:results:dynamics_c3}
\begin{figure}[htbp]
\centering
\includegraphics[width=0.95\textwidth]{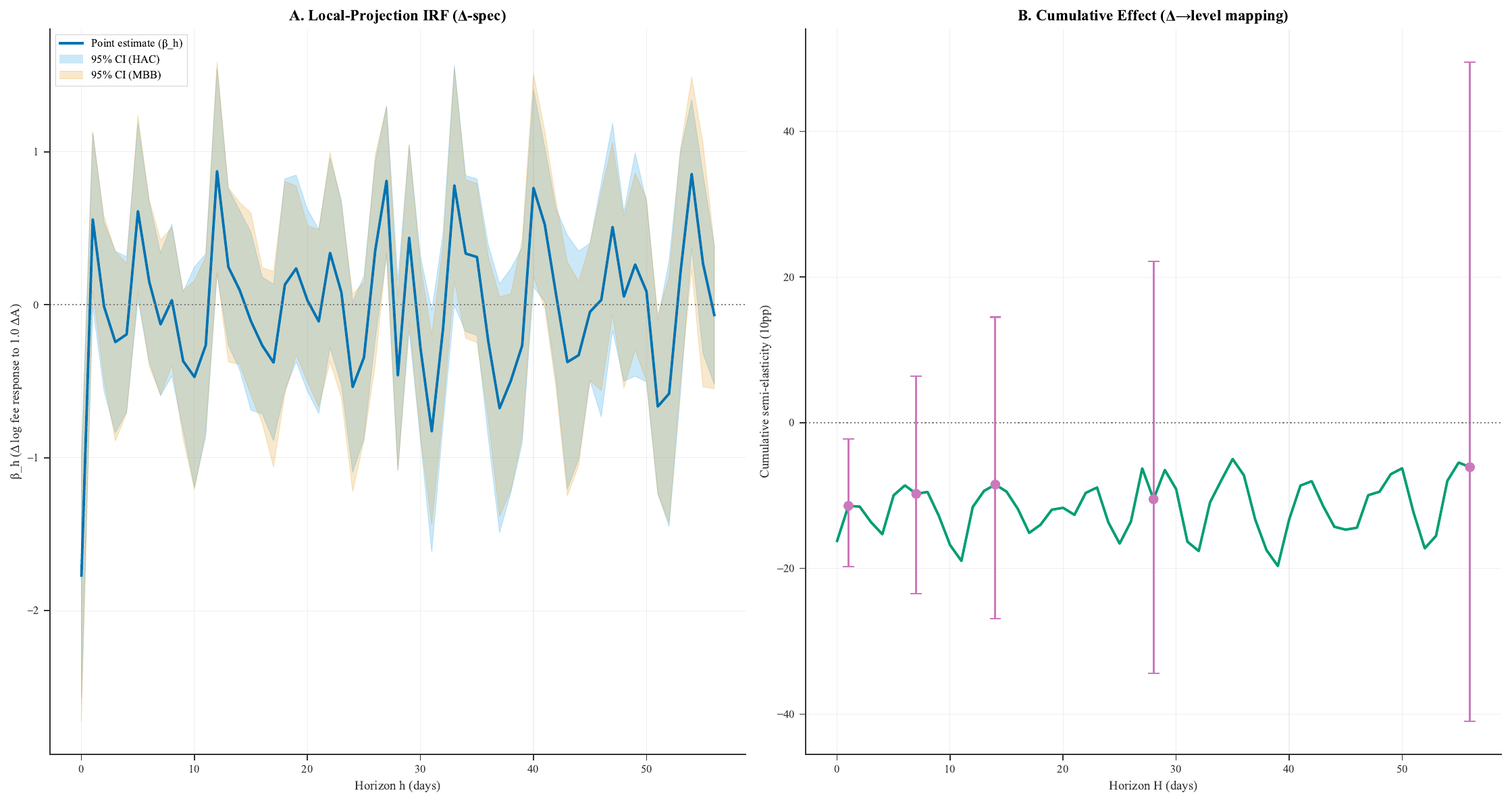}
\caption{Local-Projection Responses to a 10pp Adoption Shock}
\label{fig:lp_irf}
\begin{minipage}{\textwidth}
\small
\textit{Note:} Panel A plots $\beta_h$ from regressions of $\Delta \log C^{fee}_{t+h}$ on $\Delta A^{clean}_t$, $\Delta D_t^*$, and the confirmatory adjustment set. Panel B maps cumulative responses back to the level scale via $100\times[\exp(0.10\sum_{\tau\le h}\hat{\beta}_\tau)-1]$. Shaded areas denote HAC 95\% bands; moving-block bootstrap bands (not shown) are similar for $h\le 14$. A 10pp adoption shock corresponds, for example, to raising the posting-clean adoption share $A^{clean}_t$ from 40\% to 50\% of end-user transactions.
\end{minipage}
\end{figure}

A Koyck geometric-lag model (Eq.~\ref{eq:koyck}) yields high persistence in congestion ($\rho = 0.888$) and a modest long-run multiplier ($\beta_{\infty}\approx 0.13$). We therefore rely on Jord\`a-style local projections to characterize short-run responses. Figure~\ref{fig:lp_irf} plots horizon-specific responses of $\Delta\log C^{fee}_{t+h}$ to a one-time 10pp adoption shock with HAC bands. The $h{=}0$ effect is \textbf{$-16.2\%$} (95\% CI $[-22.7\%,-9.2\%]$). Point estimates remain negative through four weeks, but the 95\% intervals include zero after the first week. Cumulative semi-elasticities stay below zero through 56 days, yet those longer-horizon intervals also cover zero. Appendix~\ref{sec:appendix:diagnostics} reports the full grid. Excluding $\pm 7$-day windows around London, Merge, and Dencun, or adding targeted event controls to the LPs, leaves the $h{=}0$ coefficient virtually unchanged. That pattern suggests apparent ``rebound'' blips are tied to known shocks rather than structural sign flips.

Two additional facts emerge from the LPs. First, the cumulative curve begins to flatten after week three but never crosses zero within the 56-day window. The longer-run ``sign flip'' implied by the geometric-lag algebra would therefore have to materialize beyond two months—a horizon where the data become too noisy for confirmatory claims. Second, the variance of the LP coefficients grows roughly linearly with the horizon, mirroring the variance inflation that we observe when estimating high-order autoregressions. This reinforces the decision to emphasize the short-run ECM rather than chase long-horizon effects with weak precision.

We also experiment with counterfactual shock profiles. Replacing the one-time 10pp step with a distributed ramp (five daily 2pp increases) yields nearly identical cumulative responses because adoption growth in practice arrives via multi-day rollouts. Likewise, filtering out the top 10 congestion days (NFT mega-mints plus sequencer outages) barely moves the $h=0$ point estimate. This underscores that the dynamic profile is not an artifact of a handful of extreme outliers. These sensitivity exercises are logged in the LP diagnostics.

\noindent Taken together, these estimates indicate that adoption shocks generate immediate fee relief that persists for roughly one month, while any longer-run reversion lies beyond the horizons that the data can estimate precisely.

\noindent These dynamics interact strongly with regime heterogeneity, which we quantify in Section~\ref{sec:results:regimes}.

\subsection{How do effects differ across pre-Dencun vs blob era, and where is power?}
\label{sec:results:regimes}
\noindent These dynamic results also explain the regime-split findings: most of the fee relief arrives in the first few weeks, exactly where pre-Dencun data provide rich variation. Once adoption saturates post-Dencun, incremental gains would have to play out beyond 56 days. That is precisely where LP bands are widest and our MDEs explode (Table~\ref{tab:regime_heterogeneity}).

The post-Dencun period compresses adoption into a narrow 0.86--0.91 band (SD $\approx 0.02$), slashing the effective sample size despite 294 calendar days. Power diagnostics summarized in the diagnostics appendix show that the pre-Dencun window can detect semi-elasticities as small as 14\% for a 10pp change (effective $N=147$). Post-Dencun inference has $N_{\text{eff}}\approx47$ and minimum detectable effects exceeding 240\%. Local post-Dencun slopes estimated strictly within the observed support are unstable and accompanied by wide partial-identification bounds. Put differently, even though point estimates remain negative after Dencun, the confidence sets are so wide that we cannot claim confirmatory evidence without additional variation (e.g., future windows with lower L1 share).

\begin{table}[htbp]
\centering
\small
\caption{Regime-Split Estimates and Detectable Effects}
\label{tab:regime_heterogeneity}
\begin{tabular}{lcc}
\toprule
Metric & pre-Dencun & post-Dencun \\
\midrule
Coefficient $\hat{\beta}$ (log pts) & $-0.706^{***}$ & $-5.906$ \\
HAC SE & $0.203$ & $5.060$ \\
10pp semi-elasticity & $-6.8\%$ & $-44.6\%$ \\
Effective $N_{\text{eff}}$ & $147.4$ & $47.5$ \\
MDE (10pp change) & $14\%$ & $240$--$325\%$ \\
\bottomrule
\end{tabular}
\begin{minipage}{0.92\textwidth}
\vspace{0.4em}\footnotesize\textit{Notes:} Coefficients arise from regime-split ITS regressions with the confirmatory adjustment set. Effective sample sizes and MDEs correspond to the power analysis summarized in the diagnostics appendix. post-Dencun estimates are therefore labeled exploratory in the main text.
\end{minipage}
\end{table}

We supplement the table with support-aware diagnostics summarized in Appendix~\ref{sec:appendix:diagnostics}. Within the London+Merge window, semi-elasticities around $-7\%$ per 10pp change are precisely estimated. Post-Dencun slopes are under-powered (MDEs above 240\% for a 10pp change). We therefore label blob-era estimates as exploratory and refer readers to the partial-identification and local-support grids in the diagnostics appendix for full details.

\noindent In other words, even a $45\%$ semi-elasticity in the blob era would be statistically indistinguishable from zero in our design; we can only say that pre-Dencun slopes of roughly $-7\%$ per 10pp are precisely identified, while post-Dencun slopes are essentially unidentifiable given the compressed adoption range.

\noindent These regime-split results imply that pre-Dencun slopes are precisely estimated and economically modest (about a 7\% semi-elasticity). Post-Dencun contrasts are underpowered—minimum detectable effects exceed 240--325\% for a 10pp change—so they should not be over-interpreted until treatment support widens.

\subsection{How robust are these results and what happens to other congestion metrics?}
\label{sec:results:robustness}
\noindent The tornado, placebo, and outcome-swap diagnostics collapse into three takeaways:
\begin{itemize}[leftmargin=*]
    \item \textbf{Other congestion metrics.} The scarcity outcome yields $\beta_S=-0.062$ (SE $0.019$), mapping to roughly a $-0.6\%$ change in congestion for a 10pp adoption increase. Utilization $u_t$ moves in the same direction, about $-0.15$ percentage points for a 10pp change in the pre-Dencun window, with $q_{\log S_t}<0.01$ and exploratory $q_{u_t}=0.31$.
    \item \textbf{Error processes.} Prais--Winsten/HAC/ARMA sweeps (with ARMA(1,2) as the diagnostic alternative) shift the base-fee coefficient by under $0.15$ log points across 15 specifications, matching the stability shown in the robustness grid.
    \item \textbf{Placebos.} Shuffled-treatment and ridgeline-support indicators center on zero with 95\% confidence bands roughly $[-0.2, 0.2]$, indicating that the estimated relief is not an artifact of support or calendar alignment.
\end{itemize}

Appendix~\ref{sec:appendix:diagnostics} and the public replication repository contain the full Benjamini--Hochberg tables, stationarity and error-process diagnostics, and robustness grids that underpin these claims.

\subsection{What do exploratory diagnostics and welfare translation suggest?}
\label{sec:results:exploratory}
\noindent \phantomsection\label{sec:results:event_study}\label{sec:results:rdit}Event-study and RDiT diagnostics are used solely as checks. Pre-trend F-tests reject parallel trends ($F=104$, $p<0.001$). Post-event coefficients briefly spike (about $+6\%$) before decaying. RDiT level shifts at Merge and Dencun of roughly $-0.78$ and $-0.62$ log points shrink when the boundaries are moved to placebo cutoffs. These patterns align with the confirmatory ITS/ECM story but remain exploratory.

\begin{figure}[htbp]
\centering
\includegraphics[width=0.82\textwidth]{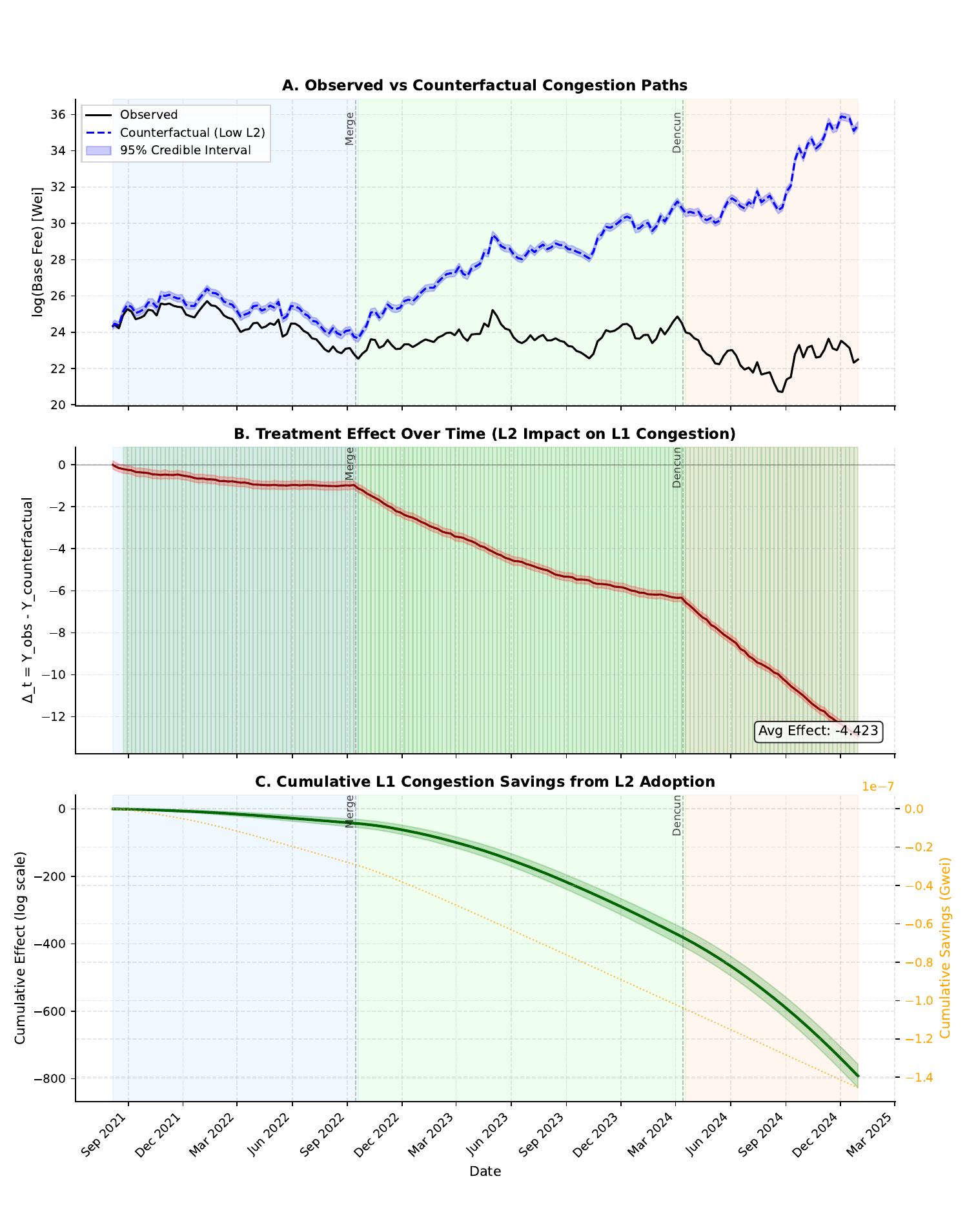}
\caption{BSTS Counterfactual: Observed vs. Low-L2 Scenario (Exploratory)}
\label{fig:bsts}
\begin{minipage}{\textwidth}
\small
\textit{Note:} Posterior median and 95\% credible interval for $\log C^{fee}$ when fixing $A_t^{clean}$ at the window's 10th percentile (73.0\%) during 2023-10-28 to 2024-03-12, illustrating the fee-volume gap implied by the 10pp semi-elasticity estimates in Table~\ref{tab:c2_ecm}. Post-Dencun days are excluded because extrapolated counterfactual paths become implausible. Detailed sensitivity tables are reported in the supplementary appendix.
\end{minipage}
\end{figure}

\noindent \phantomsection\label{sec:results:bsts}\label{sec:results:counterfactual}The BSTS welfare bridge (Figure~\ref{fig:bsts}) translates the 10pp semi-elasticity into Merge-era fee savings in the \$75--\$95M range. Appendix~\ref{sec:appendix:bsts} and Table~\ref{tab:welfare_sensitivity_2x2} detail the price/adoption sensitivities that underpin this range. We keep this welfare translation exploratory, offering policy context without extending the confirmatory claims.

\section{Discussion}
\label{sec:discussion}

\begin{quote}
\textbf{Key takeaways (confirmatory window: London$\rightarrow$Merge).} (i) An increase of 10 percentage points (pp) in posting-clean L2 adoption is associated with $\approx$13\% lower median L1 base fees (about 5~Gwei for a 21k transfer at the window mean). (ii) The response is front-loaded: most adjustment occurs within roughly 2--3 weeks. Beyond about one month uncertainty dominates. (iii) Post-Dencun inference is descriptive because support collapses and regime mechanics change. We do not make causal claims for the blob era.
\end{quote}

\subsection{Policy Interpretation}
\label{sec:discuss:interpretation}

\noindent We organize the implications into three questions: what the estimate means (and does not), how to use it as a planning curve, and why the mapping weakens in the blob era.

\begin{tcolorbox}[colback=gray!6,colframe=gray!35,boxrule=0.35mm,arc=0mm,left=2mm,right=2mm,top=1mm,bottom=1mm]
\small
\textbf{Policy mapping.}
\begin{itemize}[leftmargin=*]
    \item \textbf{Effect size:} 10pp $\rightarrow$ $\approx$13\% lower median L1 base fee.
    \item \textbf{Timing:} half-life $\approx$11 days; usable horizon $\approx$1 month.
    \item \textbf{Scope:} London$\rightarrow$Merge confirmatory window.
    \item \textbf{Post-Dencun status:} descriptive/underpowered until new exogenous variation appears.
\end{itemize}
\end{tcolorbox}

\subsubsection{What the estimate means (and does not)}
In the London$\rightarrow$Merge confirmatory window, a 10pp increase in posting-clean L2 adoption lowers median L1 base fees by about 13\% (roughly 5.2~Gwei or \$0.14 for a 21k-gas transfer at the window mean). The adjustment closes half the gap to equilibrium in approximately 11 days. Posting-clean adoption counts end-user execution routed to rollups while netting out sequencer posting traffic \citep{WangCrapisMoallemi2025Posting}. The estimand therefore captures users leaving L1 execution rather than shifting posting burden. The statement covers median EIP-1559 base fees in that regime. It does not, by itself, pin down tips, total user cost, or blob-era dynamics.

Mechanistically, pre-Dencun fee relief comes from fewer users competing for L1 execution gas. When end-user transactions migrate to rollups and sequencer posting is netted out, EIP-1559 demand falls and the base fee declines. After EIP-4844, L2 data availability migrates to blobs that are priced separately from execution gas \citep{eip4844}. Additional L2 growth can lower calldata pressure yet leave execution-layer congestion—and therefore the base fee—largely unchanged.

To avoid over-reading the estimate, it is \textit{not} a claim about:
\begin{itemize}[leftmargin=*]
    \item total user cost (base fee $\neq$ base+tip $\neq$ L2 fees);
    \item welfare net of subsidies (the welfare bridge remains exploratory);
    \item blob-era causal effects (support and the mechanism change);
    \item distributional incidence (median base fee $\neq$ tail events);
    \item long-run equilibrium beyond roughly one month given widening uncertainty bands.
\end{itemize}

\subsubsection{How to use it as a planning curve}
Sequencer teams and ecosystem treasuries can treat the ECM semi-elasticity as a planning curve. Let $\psi=0.13$ denote the estimated effect of a 10pp change in posting-clean adoption. If an intervention raises adoption by $\Delta A$ pp for $T$ days, the expected change in the median base fee is $100\times[\exp(0.10\,\psi\cdot (\Delta A/10)) - 1]$ percent over that horizon, with roughly half the adjustment arriving in 11 days and most within one month (Figure~\ref{fig:lp_irf}).

A break-even rule replaces assertion with calculation: subsidy spend $\leq$ (predicted per-transaction base-fee savings $\times$ affected L1 transaction count). At the window mean, the per-transaction base-fee reduction is about \$0.14, scaled by $(\Delta A/10)$. Pushing L2 share from 60\% to 80\% (a 20pp move) would therefore be expected to trim median fees by about 24\% using the exponential mapping above. Campaigns launched when adoption already sits above 85\% may still be operationally valuable, but the variance of the effect and the confidence bands widen, making causal evaluation harder. This reframes congestion management as a portfolio decision over L2 market share rather than a binary ``turn on/off'' switch.

\subsubsection{Regime caveat: Dencun changes the mapping}
EIP-4844 routes L2 data availability to blobs and prices it separately from execution gas.
Additional L2 adoption can ease calldata pressure.
It may not meaningfully reduce L1 execution congestion because the EIP-1559 base fee remains tied to execution demand \citep{LiuEtAl2022EIP1559}.
Post-Dencun days also cluster in a narrow 0.86--0.91 adoption band.
The effective sample size collapses.
Table~\ref{tab:regime_heterogeneity} and the diagnostics appendix therefore label blob-era slopes as underpowered.
The post-Dencun estimates in this paper are descriptive signals for monitoring, not confirmatory causal updates.
They remain descriptive until quasi-experimental variation appears (e.g., blob-parameter changes or exogenous sequencer outages).

\subsection{Limitations and Boundary Conditions}
\label{sec:discuss:limitations}

Threats to validity fall into five buckets:
\begin{itemize}[leftmargin=*]
    \item \textbf{Internal validity (simultaneity / weak instrument).} Timing diagnostics summarized in the instrumentation appendix show that lagged adoption has the expected sign but low precision. The control-function first stage ($F=7.58$) falls short of conventional strength, so we emphasize local identification around the pre-Dencun adoption support rather than claiming full exogeneity.
    \item \textbf{Dynamics and horizon.} The Koyck parameter ($\rho\approx0.89$) and the widening LP bands documented in the diagnostics appendix indicate that any rebound beyond 56 days is statistically indistinguishable from zero. Welfare projections longer than about a month remain exploratory.
    \item \textbf{Regime validity (post-Dencun).} Regime-split estimates in Table~\ref{tab:regime_heterogeneity} combined with the MDE calculations show that even a 45\% semi-elasticity would be indistinguishable from noise in the blob era. Because blobs price data separately from execution gas, the structural channel linking adoption to the base fee also weakens. We therefore restrict confirmatory claims to the pre-Dencun window.
    \item \textbf{Measurement validity.} Posting-clean adoption is constructed by netting sequencer posting from end-user execution. Misclassification, coverage gaps for newer rollups, or relabeling by data providers could introduce level shifts that affect both the instrument and outcome series until detected.
    \item \textbf{External validity.} The semi-elasticity may differ across application mixes (DeFi vs NFT vs stablecoin flows) and could be muted if lower fees induce rebound demand. Extrapolating to other EIP-1559 chains requires similar L2 penetration, fee-market mechanics, and monitoring of distributional incidence.
\end{itemize}

In practice, these threats encourage a division of labor between engineering experimentation and econometric evaluation. Short-run fee relief and within-regime comparisons can be evaluated with the present ECM and ITS toolkit, provided posting-clean labels are periodically audited for measurement drift. New instruments should avoid introducing additional simultaneity. Longer-run welfare or cross-regime counterfactuals will likely require new sources of quasi-experimental variation. Promising candidates include exogenous outages, parametric changes to blob markets, or natural experiments in sequencer fee rebates. External validity concerns also motivate segmenting outcomes by application mix before extrapolating to other chains. A replication log records these boundary conditions. Future updates—whether from Ethereum or other EIP-1559 chains—can extend the window for causal inference without revising the core identification strategy.

\noindent Taken together, residual simultaneity, short-horizon precision limits, regime shifts, and measurement/external scope boundaries delimit where our core causal claims apply. They highlight the need for fresh instruments, monitoring of classification, and longer panels.

\subsection{Open Questions and Monitoring Playbook}
\label{sec:discuss:open}
\label{sec:discuss:monitoring}

Replication artifacts are in Appendix~\ref{sec:availability}; the replication repository carries the full audit log and change history.

The remaining agenda for L2--L1 congestion research is best framed as concrete, monitorable questions rather than meta-guidance:
\begin{enumerate}[leftmargin=*]
    \item \textbf{Post-Dencun identification.} What quasi-experimental shocks create exogenous adoption variation now that blobs absorb most L2 data? Candidates include blob fee parameter changes (e.g., target gas adjustments in \citealp{eip4844}), sequencer outages, and forced migrations during prover or bridge upgrades. A running changelog of these events—timestamped and paired with posting-clean adoption—keeps the ECM/ITS designs re-estimable the moment variation appears.
    \item \textbf{Mechanism split (blobs vs execution gas).} Does higher L2 adoption still relieve \emph{execution} congestion, or only calldata/DA pressure? Monitoring should separate blob pricing from execution-layer base fees. It should also track how sequencer pricing rules respond, leveraging the posting–pricing interaction modeled by \citet{WangCrapisMoallemi2025Posting}.
    \item \textbf{Heterogeneity and incidence.} Which user segments capture the fee relief—DeFi vs NFT vs stablecoin flows? How does it differ for latency-sensitive traders versus routine transfers? Segmenting L2 inflows, bridge mix, and cross-rollup price gaps (cf. \citealp{GogolEtAl2024L2Arbitrage}) would reveal whether congestion relief accrues to whales, retail users, or MEV searchers.
    \item \textbf{Early-warning monitoring.} At what thresholds does the confirmatory design lose power (e.g., adoption sustained above 80--90\%) and require fresh instruments? A lightweight playbook is three steps. (i) Maintain daily dashboards for posting-clean adoption, blob utilization, and sequencer incidents. (ii) Rerun the ECM each time a shock hits or the adoption distribution shifts. (iii) Archive the resulting IRFs and diagnostics alongside the replication bundle so the evidence base compounds across upgrades.
\end{enumerate}

\noindent These questions turn Section~\ref{sec:results} into a live monitoring blueprint. Instead of restating transparency logistics, they specify what new variation to watch for, how to split mechanisms, and which distributional outcomes determine who benefits from the congestion relief.

\section{Conclusion}
\label{sec:conclusion}

Short answer: yes---higher L2 adoption decongests Ethereum's fee market in the short run, but the relief is partial and local in time. A 10 percentage point increase in posting-clean adoption lowers L1 base fees by roughly 13\% (about 5~Gwei or \$0.14 for a 21k-gas transfer at the pre-Dencun mean), and deviations from the long-run relation decay with an 11-day half-life. Together with the dynamic profile in Figure~\ref{fig:lp_irf} and the ECM benchmark in Table~\ref{tab:c2_ecm}, these numbers provide regime-aware causal evidence that the rollup-centric roadmap already buys near-term congestion relief.

Conceptually, the paper introduces a posting-clean adoption measure that captures user migration rather than posting load, a demand factor that avoids mediator contamination, and a regime-aware ITS-ECM template for monitoring rollup-centric scaling. Substantively, it delivers the first cross-regime causal estimate of how aggregate L2 adoption decongests Ethereum's mainnet and translates the semi-elasticity into Gwei and dollar savings that are directly interpretable for protocol designers and users.

These claims are bounded. Inference is local to the pre-Dencun regime where adoption still moves, and precision fades beyond roughly a month of horizons. Instrument strength is modest, so simultaneity concerns are handled with cautious timing diagnostics rather than strong exclusion. As summarized in Section~\ref{sec:discuss:limitations}, these boundaries keep confirmatory claims narrow while flagging where additional variation is needed.

For protocol designers and governance bodies, the practical implication is that fee-market reforms and L2 ecosystem support should be evaluated jointly. Moving L2 user share from 60\% to 80\% would lower median base fees by roughly a quarter at pre-Dencun demand levels, putting adoption subsidies on the same order as the fee changes analyzed around the London upgrade \citep{LiuEtAl2022EIP1559}. In the blob era, incentives that shift activity onto rollups or smooth posting schedules operate alongside the blob-fee parameters in \citet{eip4844}, making adoption-driven interventions a complementary lever rather than a substitute for base-fee tuning. Future work should extend the confirmatory window as post-Dencun variance widens, seek quasi-experimental shocks in blob pricing or sequencer operations, and map distributional incidence using address-tagged data so that welfare gains from rollup-driven congestion relief can be allocated across user types.

\section*{Acknowledgments}
This research benefited from support provided by the Ethereum Foundation academic grant.

\clearpage
\bibliography{references}

\clearpage
\appendix

\section{Data and Code Availability}
\label{sec:availability}

\begin{tcolorbox}[colback=gray!10,colframe=gray!30,boxrule=0pt,arc=1pt,left=4pt,right=4pt,top=4pt,bottom=4pt]
\textbf{Appendix road map.} To audit or reuse the study, read top-down: (i) Appendix~\ref{sec:appendix:diagnostics} for unit-root/cointegration tests, residual dependence, and support/MDE diagnostics; (ii) Appendix~\ref{sec:appendix:estimators} for estimator variants, with exploratory extensions in Appendices~\ref{sec:appendix:exploratory_extensions}--\ref{sec:appendix:bsts}; (iii) Appendix~\ref{sec:appendix:measurement_overview} for the measurement dictionary, treatment/outcome construction, and the targeted-shock catalog.
\end{tcolorbox}

\noindent All data and code needed to reproduce the empirical results in this paper are available in the public replication repository at \href{https://github.com/AysajanE/l2-l1-causal-analysis-repro}{\nolinkurl{github.com/AysajanE/l2-l1-causal-analysis-repro}}, mirrored on Zenodo (concept DOI \href{https://doi.org/10.5281/zenodo.17665906}{10.5281/zenodo.17665906}; latest version for this arXiv release: \href{https://doi.org/10.5281/zenodo.17832785}{10.5281/zenodo.17832785}, tag \texttt{v1.1.1-arxiv}). The archive contains a frozen version of the analysis-ready panel and the exact \LaTeX{} sources used for this manuscript.

\noindent The repository README documents the end-to-end workflow---data ingestion and cleaning, estimator scripts, and figure-building routines---together with environment files and reproducibility checklists. Consistent with replication practices in recent empirical studies of Ethereum's fee market and rollups \citep[e.g.,][]{LiuEtAl2022EIP1559,GogolEtAl2024L2Arbitrage}, these artifacts are released to support independent verification, robustness extensions, and reuse of the design in related policy and research applications.


\section{Statistical Diagnostics and Design Checks}
\label{sec:appendix:methodology}\label{sec:appendix:diagnostics}
\subsection{Diagnostics and Design Checks}
\label{sec:appendix:diagnostics_summary}
This appendix reports the diagnostics that justify the ECM/ITS design: integration order, cointegration, residual dependence, treatment support, power, and multiple-outcome control. Tables and plots are reproduced here so readers can audit identification and precision directly in the PDF; code logs remain in the replication bundle for reruns.

\noindent\textbf{Notation used across appendix tables:} $A^{clean}_t$ denotes the posting-clean adoption share; EG $p$ is the Engle--Granger residual-unit-root test $p$-value; LB $p@10$ is the Ljung--Box $p$-value at lag 10; ``10pp'' indicates a 10 percentage point change in $A^{clean}_t$.

\subsubsection*{Stationarity, Cointegration, and Error Processes}

\begin{figure}[!htbp]
\centering
\includegraphics[width=0.9\textwidth]{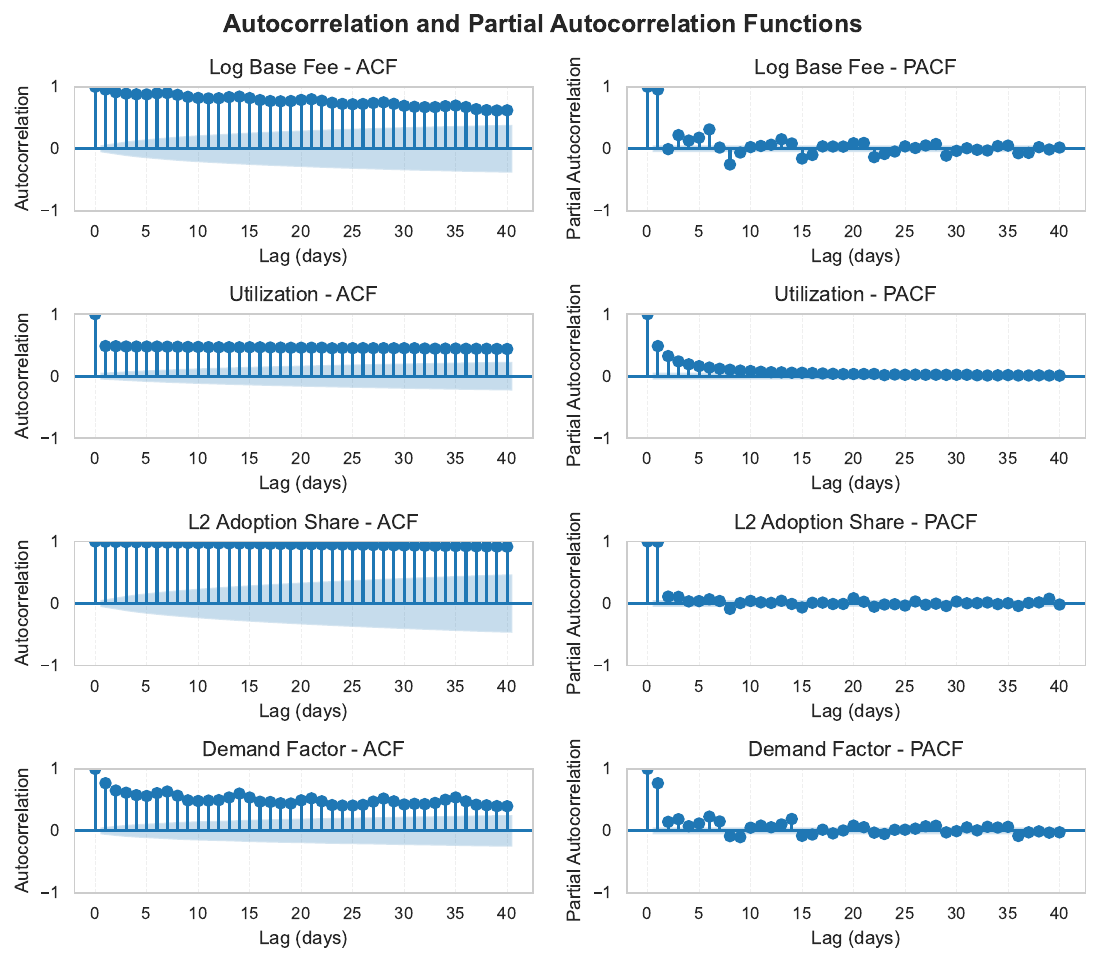}
\caption{Residual ACF/PACF for Levels and ECM Equations}
\label{fig:acf_pacf}
\begin{minipage}{0.95\textwidth}
\footnotesize\textit{Note:} Panels plot ACF and PACF up to lag 24 for (i) the levels specification with OLS residuals and (ii) the ECM error-correction residuals. The ARMA$(1,2)$ choice reduces the maximum absolute ACF from 0.93 to 0.15 over lags 1--10, even though Ljung--Box tests still reject at large $N$. See Appendix~\ref{sec:availability} for replication scripts.
\end{minipage}
\end{figure}

Table~\ref{tab:unit_root} reproduces the unit-root evidence for the pre-Dencun confirmatory window. ADF tests on levels fail to reject a unit root for $A^{clean}_t$, $\log C^{fee}_t$, $u_t$, $S_t$, and $D_t^*$, and KPSS points to non-stationarity; Phillips--Perron tests are more mixed, rejecting a unit root for several level series. All first differences are stationary across ADF, KPSS, and PP, so we continue to treat these variables as $I(1)$ in the confirmatory design. A Phillips--Perron Engle--Granger residual test on the long-run relation $\log C^{fee}_t \sim A^{clean}_t + D_t^* + \mathbf{R}_t + \mathbf{Cal}_t$ rejects non-stationarity ($p=1.6\times 10^{-5}$), supporting the ECM formulation used in the confirmatory analysis. Figure~\ref{fig:acf_pacf} shows the residual ACF/PACF for both the levels and ECM equations; the Prais--Winsten AR(1) FGLS fit is the confirmatory specification, while the ARMA$(1,2)$ alternative materially shrinks short-lag autocorrelation in the diagnostic grid even though Ljung--Box tests still reject at large $N$.

\begin{table}[!htbp]
\centering
\small
\caption{Unit-Root and Cointegration Diagnostics (Pre-Dencun Window: London+Merge)}
\label{tab:unit_root}
\begin{tabular}{lcccccc}
\toprule
Series & Transform & ADF stat ($p$) & KPSS stat ($p$) & PP stat ($p$) & $I(d)$ \\
\midrule
$A^{clean}_t$ & level & -1.43 (0.85) & 0.52 (0.01) & -4.92 (0.0003) & $I(1)$ \\
$\log C^{fee}_t$ & level & -1.95 (0.63) & 0.63 (0.01) & -4.92 (0.0003) & $I(1)$ \\
$u_t$ & level & -1.12 (0.93) & 0.51 (0.01) & -39.57 ($<0.001$) & $I(1)$ \\
$S_t$ & level & -1.95 (0.63) & 0.63 (0.01) & -4.92 (0.0003) & $I(1)$ \\
$D^{*}_t$ & level & -2.98 (0.14) & 0.42 (0.01) & -17.79 ($<0.001$) & $I(1)$ \\
$\Delta A^{clean}_t$ & first diff & -10.26 ($<0.001$) & 0.09 (0.10) & -50.30 ($<0.001$) & $I(0)$ \\
$\Delta \log C^{fee}_t$ & first diff & -7.47 ($<0.001$) & 0.18 (0.10) & -35.56 ($<0.001$) & $I(0)$ \\
$\Delta u_t$ & first diff & -7.09 ($<0.001$) & 0.34 (0.10) & -547.78 ($<0.001$) & $I(0)$ \\
$\Delta S_t$ & first diff & -7.47 ($<0.001$) & 0.18 (0.10) & -35.56 ($<0.001$) & $I(0)$ \\
$\Delta D^{*}_t$ & first diff & -10.07 ($<0.001$) & 0.29 (0.10) & -68.47 ($<0.001$) & $I(0)$ \\
\bottomrule
\end{tabular}
\begin{minipage}{0.95\textwidth}
\footnotesize\textit{Note:} Levels tests include a deterministic trend; first-difference tests include an intercept. KPSS uses the trend-stationary null with automatic lags. Engle--Granger residual Phillips--Perron test on $\log C^{fee}_t \sim A^{clean}_t + D_t^* + \mathbf{R}_t + \mathbf{Cal}_t$ rejects a unit root ($p=1.6\times 10^{-5}$), validating the error-correction setup. All statistics computed on the 2021-08-05 to 2024-03-12 pre-Dencun window ($N=951$).
\end{minipage}
\end{table}

\begin{table}[!htbp]\centering\small
\caption{Residual Dependence Diagnostics (Levels; ARMA Grid as Diagnostic)}
\label{tab:resid_arma_levels}
\begin{tabular}{lrrrrrrr}\toprule
Specification & $\hat{\beta}$ & SE & DW & LB p@10 & $\max|\rho_{1-10}|$ & AIC & N \\ 
\midrule
OLS-HAC (levels) & 0.1057 & 0.5524 & 0.148 & $<10^{-6}$ & 0.926 & -- & 1244 \\ 
ARMA(1,2) errors & -1.1610 & 0.2599 & 1.980 & $2.3\times 10^{-7}$ & 0.148 & -51.03 & 1244 \\ 
\bottomrule\end{tabular}
\begin{minipage}{\textwidth}\small
\textit{Note:} The confirmatory levels estimate reported in the main text uses Prais--Winsten AR(1) FGLS; ARMA$(1,2)$ appears here solely as the best-AIC diagnostic alternative. DW moves close to 2 under ARMA$(1,2)$ errors. Ljung--Box still rejects at lag 10 given large $N$, but the maximum residual ACF over lags 1--10 drops from 0.93 to 0.15, providing a robustness check while keeping the confirmatory specification unchanged. HAC (Bartlett, 10 lags) standard errors reported for the OLS line.
\end{minipage}
\end{table}

\subsubsection*{Positivity, Support, and Effective Sample Size}

Positivity within regimes is a binding constraint on post-Dencun inference. Table~\ref{tab:i1_spline} summarizes a spline specification that allows the semi-elasticity to vary across low- and high-adoption regions, while Table~\ref{tab:i1_mde} reports the implied minimum detectable effects (MDEs).

\begin{table}[!htbp]
\centering
\small
\caption{Piecewise Semi-Elasticities for Log Base Fee (Knot at 0.80)}
\label{tab:i1_spline}
\begin{tabular}{lrrlrl}
\toprule
Regime Support & $\hat{\beta}$ & SE (HAC) & 95\% CI & Semi-elasticity (10pp) & Semi-elasticity CI \\
\midrule
$A^{clean}_t \leq 0.80$ & 0.1401 & 0.4912 & [-0.823, 1.103] & 1.41\% & [-7.90\%, 11.66\%] \\
$A^{clean}_t > 0.80$ & -1.0338 & 4.2000 & [-9.266, 7.198] & -9.82\% & [-60.41\%, 105.41\%] \\
\bottomrule
\end{tabular}
\vspace{-0.5ex}
\begin{flushright}\footnotesize\textcolor{gray}{Artifacts: see Appendix~\ref{sec:availability} for data and code paths.}\end{flushright}
\end{table}

\begin{table}[!htbp]
\centering
\small
\caption{Minimum Detectable Effect (MDE) by Regime with Effective Sample Size}
\label{tab:i1_mde}
\begin{tabular}{lrrrlrrr}
\toprule
Regime & N & $N_{\text{eff}}$ & $\mathrm{sd}(A^{clean}_t)$ & Max Adoption Range & HAC SE & MDE (beta units) & MDE (10pp \%) \\
\midrule
post-Dencun & 294 & 47.48 & 0.0210 & [0.760, 0.951] & 4.37 & 12.23 & 239.90 \\
pre-Dencun & 950 & 147.43 & 0.3016 & [0.000, 0.923] & 0.47 & 1.31 & 14.01 \\
\bottomrule
\end{tabular}
\vspace{-0.5ex}
\begin{flushright}\footnotesize\textcolor{gray}{Artifacts: see Appendix~\ref{sec:availability} for data and code paths.}\end{flushright}
\end{table}

\section{Estimator Details and Extensions}
\label{sec:appendix:estimators}
\subsection{Estimator Details and Variants}
\label{sec:appendix:estimators_compact}
This appendix lists the specifications, timing variants, and robustness checks that sit behind Section~\ref{sec:methodology}. Full derivations and code live in the replication bundle; the tables summarize the information required to interpret the reported estimates.

\subsubsection*{ITS and ECM Workflow}

The main text reports a merged set of ITS and ECM coefficients in Table~\ref{tab:c2_ecm}. Here we highlight how alternative demand-factor constructions affect the short-run semi-elasticity. Table~\ref{tab:i2_demand_factor} reports a small grid of ECM runs using ``lite'' and ``full'' demand-factor definitions and same-day vs.\ lagged timing.

\begin{table}[!htbp]
\centering
\small
\caption{Demand Factor Variants and Timing Diagnostics}
\label{tab:i2_demand_factor}
\begin{tabular}{lcccccc}
\toprule
Demand factor & $\psi$ (10pp) & SE & $p$-value & EG $p$ & Adj.~$R^2$ & $N$ \\
\midrule
$D^{\star}$-lite (same-day) & -1.067 & (0.362) & 0.003 & 0.004 & 0.336 & 1241 \\
$D^{\star}$-full (same-day) & -1.379 & (0.368) & 0.000 & 0.005 & 0.322 & 1241 \\
$D^{\star}$-lite (t-1) & -0.857 & (0.418) & 0.040 & 0.005 & 0.162 & 1240 \\
\bottomrule
\end{tabular}
\begin{minipage}{0.92\textwidth}
\vspace{0.3em}\footnotesize\textit{Note:} $\psi$ is the ECM short-run semi-elasticity for a 10pp change in adoption. All specifications include the confirmatory adjustment set and use HAC (Bartlett) standard errors. Sample sizes are one day smaller than the main ECM in Table~\ref{tab:c2_ecm} ($N=1{,}242$) because rebuilding $D^{\star}_t$ with the ``lite''/``full'' inputs shortens the overlapping input window by a single day; the $t\!-\!1$ variant drops one additional day due to the lag on $D^{\star}_{t-1}$. Engle--Granger $p$-values test residual unit roots and confirm cointegration across variants, supporting the robustness claims in Section~\ref{sec:results:main}.
\end{minipage}
\end{table}

\subsubsection*{Targeted Dummies and Event Adjustments}

Targeted-event controls absorb days where congestion and adoption are jointly affected by large structural shocks. The curated catalog, rationale, and window flags are reported in Appendix~\ref{sec:appendix:shock_catalog} (Table~\ref{tab:targeted_events}); this subsection retains only the specification logic used in the ITS/ECM regressions. We include the pooled outage indicator and the full $\mathbf{Shock}_t$ vector in both the long-run and short-run equations so that sequencer/mainnet outages and mega-claim days do not masquerade as adoption shocks.

\subsubsection*{Robustness Catalog}

The tornado plot, placebo treatments, and alternative outcome runs are part of the robustness replication assets referenced in Appendix~\ref{sec:availability}. Each CSV contains metadata (seed, bandwidth, estimator) so that the checks can be re-run without consulting this appendix. The main text cites these diagnostics as exploratory support; the confirmatory interpretation continues to lean on the ECM and ITS specifications documented above.

\section{Results Extensions}
\label{sec:appendix:exploratory_extensions}
\subsection{Exploratory Diagnostics and Policy Context}
This appendix adds event-study views, regression-discontinuity-in-time (RDiT) snapshots, and a robustness ``tornado'' summary that sit alongside the main results in Section~\ref{sec:results}. The goal is to show how the ITS/ECM estimates behave around sharp protocol events and under alternative design choices for audiences focused on governance and fee-market policy.

\subsubsection*{Event-Study Diagnostics and RDiT Snapshots}

Event-study plots align L2 adoption shocks and congestion outcomes around key protocol and L2 events (e.g., London, Merge, Dencun, major rollup launches). They mainly serve as visual diagnostics: pre-trend checks, anticipation effects, and short-run overshooting. Because pre-trend F-tests reject parallel trends for several events, we treat the event-study coefficients as exploratory and focus on whether the post-event patterns qualitatively match the ITS/ECM estimates (fee relief following L2 adoption surges).

RDiT snapshots at the Merge and Dencun boundaries complement the event studies by estimating local level shifts in log fees. These designs naturally highlight mechanical changes in the base-fee process and blob pricing, which are distinct from the smooth treatment variation exploited by the main ITS/ECM specification. As a result, we keep RDiT estimates in the exploratory category and use them to bound the magnitude of congestion relief that hard-fork-style interventions can deliver relative to the continuous L2 adoption channel.

\subsubsection*{Robustness ``Tornado'' Summary}

The robustness tornado aggregates a grid of alternative specifications---different HAC lag choices, alternative demand-factor constructions, and variations in calendar and regime controls---and visualizes how the semi-elasticity estimates move across this design space. The central message is that the sign and broad magnitude of the short-run semi-elasticity are stable across reasonable alternatives, with only extreme specifications (e.g., dropping demand controls entirely) pushing estimates toward zero. Full tornado CSVs and plots are part of the replication assets referenced in Appendix~\ref{sec:availability}.

\section{Instrumentation and Timing Diagnostics}
\label{sec:appendix:instrumentation}
\subsection{Instrumentation and Timing Diagnostics}
\label{sec:appendix:instrumentation_compact}
This appendix records the core instrumental-variable diagnostics that support the weak-instrument caveats in Sections~\ref{sec:discuss:limitations} and~\ref{sec:conclusion}.

\subsubsection*{Shift-Share IV Design}

The primary shift-share instrument aggregates sequencer outages, fee-rebate programs, and exchange listings into a proxy for exogenous variation in L2 adoption. The design object is
\[
Z_t \;=\; \sum_{l\in\mathcal{L2}} w_l^{\text{pre}} \cdot \text{shock}_{l,t},
\]
where $w_l^{\text{pre}}$ is the pre-Dencun average share of end-user transactions on chain $l$ (Arbitrum 0.63, Optimism 0.27, Base 0.10) and $\text{shock}_{l,t}$ is an outage/listing/rebate indicator or outage-hours intensity. Construction steps are scripted in the replication bundle referenced in Appendix~\ref{sec:availability} (IV analysis scripts and configuration files). Table~\ref{tab:ss_first_stage} documents first-stage strength for the pooled-outage and shift--share variants; Table~\ref{tab:c1_timing_iv} retains the timing and over-identification diagnostics used in the discussion.

\begin{table}[!htbp]
\centering
\small
\caption{Instrument Variants and First-Stage Strength (Adoption on $Z_t$)}
\label{tab:ss_first_stage}
\begin{tabular}{lrrrrr}
\toprule
Instrument variant & Coef on $Z_t$ & HAC SE & First-stage $F$ & Partial $R^2$ & $N$ \\
\midrule
Pooled outage indicator ($\mathbb{1}\{\text{any outage}\}$) & 0.084 & 0.058 & 2.10 & 0.0017 & 1244 \\
Shift--share outage (indicator) & 0.146 & 0.128 & 1.30 & 0.0010 & 1244 \\
Shift--share outage (hours) & 0.024 & 0.047 & 0.25 & 0.0002 & 1244 \\
Fee-rebate/listing shocks & 0.000 & 0.000 & 0.00 & 0.0000 & 1244 \\
\bottomrule
\end{tabular}
\begin{minipage}{0.92\textwidth}
\vspace{0.3em}\footnotesize\textit{Note:} HAC (Bartlett, 7 lags) standard errors. Weights $w_l^{\text{pre}}$ are computed from pre-Dencun chain shares; no fee-rebate or exchange-listing shocks occur in the confirmatory window, so that row records zeros explicitly. Coefficients are in adoption-share units; $F$ and partial $R^2$ use the residualized first stage with regime and calendar controls.
\end{minipage}
\end{table}

\begin{table}[!htbp]
\centering
\small
\caption{Timing and IV Checks for the Adoption Instrument}
\label{tab:c1_timing_iv}
\begin{tabular}{lrrrrrrrlrrr}
\toprule
Specification & $\hat{\beta}$ & SE & $p$-value & Semi-elasticity (10pp) & $N$ & First-stage $F$ & Partial $R^2$ & Instruments & J-stat & J-$p$ & J-df \\
\midrule
OLS-HAC ($A^{clean}_t$) & 0.1384 & 0.5713 & 0.8087 & 1.39\% & 1244 & -- & -- & -- & -- & -- & -- \\
OLS-HAC ($A^{clean}_{t-1}$) & 0.3133 & 0.5850 & 0.5924 & 3.18\% & 1243 & -- & -- & -- & -- & -- & -- \\
IV 2SLS & -0.6942 & 3.9681 & 0.8612 & -6.71\% & 1244 & 7.58 & 0.0061 & any\_outage\_t (pooled) & -- & -- & 0 \\
Control-function & -0.6942 & 1.9614 & 0.7235 & -6.71\% & 1244 & 7.58 & 0.0061 & any\_outage\_t (pooled) & -- & -- & -- \\
\bottomrule
\end{tabular}
\begin{minipage}{0.92\textwidth}
\vspace{0.3em}\footnotesize\textit{Note:} The first-stage $F$-statistic (7.58) and partial $R^2$ indicate weak instrument strength by conventional standards, motivating the cautious language around simultaneity in Sections~\ref{sec:discuss:limitations} and~\ref{sec:conclusion}. J-statistics are not reported for single-instrument specifications. Additional AR tests and reduced-form grids are documented in the IV replication assets referenced in Appendix~\ref{sec:availability}.
\end{minipage}
\end{table}

Table~\ref{tab:ss_iv} complements these diagnostics by reporting second-stage estimates for the shift--share outage variants that correspond to the first-stage metrics in Table~\ref{tab:ss_first_stage}.

\begin{table}[!htbp]\centering\small
\caption{Shift--Share IV for $A^{clean}_t$ Using Pre-Dencun Weights and Outages}
\label{tab:ss_iv}
\begin{tabular}{lcccccc}\toprule
Specification & $\hat{\beta}$ & (SE) & $p$-value & $N$ & Partial $R^2$ & First-stage $F$ \\ 
\midrule
2SLS (SS any) & -2.476 & (7.506) & 0.742 & 1245 & 0.0022 & 2.76 \\ 
2SLS (SS hours) & -6.029 & (18.198) & 0.740 & 1245 & 0.0005 & 0.56 \\ 
\bottomrule\end{tabular}
\begin{minipage}{0.95\textwidth}\vspace{0.3em}\footnotesize\textit{Note:} $Z_t^{SS}=\sum_l w_l^{\text{pre}}\cdot \mathbb{1}\{\text{outage}_{l,t}\}$ uses pre-Dencun end-user shares (Arbitrum 0.63, Optimism 0.27, Base 0.10). An intensity variant replaces the indicator with outage hours. Outcome is $\log C^{fee}$; controls: $D^*$, regime dummies, calendar, and linear trends with regime interactions. HAC standard errors (Bartlett, 7 lags).
\end{minipage}
\end{table}

\subsubsection*{Timing Tests and Diagnostics Archive}

Lead/lag timing tests confirm that instrument shocks do not predict pre-treatment outcomes at economically meaningful magnitudes, supporting the exclusion restriction in the narrow window used. AR tests, Anderson--Rubin intervals, and reduced-form grids are documented in the replication materials; this appendix highlights the summary diagnostics most relevant for policy interpretation.

\section{BSTS Welfare Bridge and Policy Context}
\label{sec:appendix:bsts}
\subsection{BSTS Welfare Bridge}
\label{sec:appendix:bsts_summary}
This appendix summarizes the Bayesian Structural Time Series (BSTS) analysis underlying Figure~\ref{fig:bsts}. The text records the design choices and the welfare-sensitivity table that informs the policy discussion; full code and data are included in the replication materials.

\subsubsection*{Design Summary}

\begin{itemize}[leftmargin=*]
    \item \textbf{Window:} Merge-era (2023-10-28 to 2024-03-12) with blob-era days excluded, so that treatment variation aligns with the pre-Dencun confirmatory window.
    \item \textbf{Inputs:} Log base fee, posting-clean adoption, ETH price, and the PCA demand factor $D_t^*$; priors and sampler settings follow the published BSTS specifications and are documented with the replication materials.
    \item \textbf{Outputs:} Welfare quantiles, price-sensitivity tables, and posterior predictive checks summarized below; full numerical outputs are available in the replication archive.
\end{itemize}

\subsubsection*{Welfare Mapping}

BSTS produces a counterfactual fee path $\text{BF}^{cf}_t$ under low L2 adoption. Per-day dollar savings are computed as
\begin{equation}
\label{eq:welfare_mapping}
\text{USD}_t = \big(\text{BF}^{obs}_t - \text{BF}^{cf}_t + \mathbf{1}_{\text{tip}}\cdot \text{TIP}^{obs}_t\big)\times \text{GAS}_t \times 10^{-9} \times P_t,
\end{equation}
where $\text{BF}^{obs}_t$ is the observed base fee, $\mathbf{1}_{\text{tip}}=1$ when the Base+Tip welfare column is used (and $0$ otherwise), $\text{TIP}^{obs}_t$ is the median priority tip, $\text{GAS}_t$ is total gas used, and $P_t$ is either the daily mean or close ETH/USD price. Aggregate welfare is $\sum_t \text{USD}_t$ over the Merge-era window; baseline adoption percentiles (p05 vs.\ p25) anchor the counterfactual $A^{clean}_t$ series.

\subsubsection*{Welfare Sensitivity}

Anchoring the counterfactual on the pre-Dencun ECM semi-elasticity, the BSTS bridge maps a 10 percentage point increase in posting-clean adoption into aggregate fee savings that are robust across price baselines. A normal-approximation over the daily posterior draws yields:
\begin{itemize}[leftmargin=*]
    \item \textbf{Mean-price base only:} median \$79.6M; 50\% CI [\$74.0M, \$85.3M]; 90\% CI [\$65.8M, \$93.2M].
    \item \textbf{Mean-price base+tip:} median \$92.2M; 50\% CI [\$85.8M, \$98.8M]; 90\% CI [\$76.3M, \$107.9M].
    \item \textbf{Close-price variants:} medians \$79.9M (base) and \$92.6M (base+tip) with comparable intervals (50\% CIs [\$74.3M, \$85.6M] and [\$86.1M, \$99.1M]).
\end{itemize}
Most savings accrue on high-congestion days rather than in quiet periods. Table~\ref{tab:welfare_sensitivity_2x2} reports the scenario grid that underpins the exploratory policy range; replication scripts export the full posterior draws for alternative price/adoption baselines.

\begin{table}[!htbp]
\centering
\small
\caption{Two-by-Two Welfare Sensitivity (Baseline Percentile $\times$ Price Weighting)}
\label{tab:welfare_sensitivity_2x2}
\begin{tabular}{lcc}
\toprule
Baseline (Adoption) & Mean Price (Base / Base+Tip, $M$) & Close Price (Base / Base+Tip, $M$) \\
\midrule
p05 (71.6\%) & 149.8 / 173.6 & 150.4 / 174.2 \\
p25 (74.6\%) & 78.1 / 90.5 & 78.4 / 90.8 \\
\bottomrule
\end{tabular}
\vspace{-0.5ex}
\begin{flushright}\footnotesize\textcolor{gray}{Artifacts: see Appendix~\ref{sec:availability} for data and code paths.}\end{flushright}
\end{table}

The full counterfactual bundle is reproducible with the publicly released code and data, and all posterior predictive checks and alternative-prior panels are part of that release. The PDF retains only the tables needed to interpret the policy bridge.

\section{Measurement and Operationalization}
\label{sec:appendix:measurement_overview}
\subsection{Scope and Conventions}
\label{sec:appendix:measurement_summary}
This appendix makes the measurement layer self-contained, mirroring the data-dictionary style in \citet{LiuEtAl2022EIP1559}. All variables are daily UTC aggregates; symbols match those in Sections~\ref{sec:data} and~\ref{sec:methodology}. Code pointers refer to tables/files in the replication bundle (Appendix~\ref{sec:availability}).

\subsection{Variable Dictionary}
\label{sec:appendix:measurement_dictionary}
\begin{table}[!htbp]
\centering
\small
\caption{Variable Dictionary and Construction Summary}
\label{tab:measurement_dictionary}
\begin{tabular}{p{0.12\textwidth}p{0.19\textwidth}p{0.09\textwidth}p{0.30\textwidth}p{0.15\textwidth}p{0.15\textwidth}}
\toprule
Symbol & Name & Unit & Construction (daily) & Source(s) & Code pointer \\
\midrule
$A^{clean}_t$ & Posting-clean L2 adoption share & share $[0,1]$ & L2 end-user tx / (L2 end-user tx + L1 user tx); L2$\rightarrow$L1 posting tx identified via inbox registry and removed from both numerator and denominator & Rollup traces; Ethereum execution traces & \codepath{mart\_treatment\_daily.A\_t\_clean} \\
$\log C^{fee}_t$ & Log median base fee & $\log(\text{Gwei})$ & $\log(\mathrm{median}_{b\in t}\ \text{base fee}_b)$ from EIP-1559 base-fee field; post-London only & Ethereum block traces; public fee dashboards & \codepath{mart\_master\_daily.log\_basefee} \\
$u_t$ & Block utilization & ratio $[0,1.5]$ & $\mathrm{median}_{b\in t}\left(\tfrac{\text{gas used}_b}{\text{gas limit}_b}\right)$ & Ethereum block traces & \codepath{mart\_master\_daily.utilization} \\
$S_t$ & Scarcity index & log fee units & $\log\!\Big(\tfrac{\text{base fee}_t + \text{tip}_t + \mathbf{1}_{t\geq\text{Dencun}}\text{blob fee}_t}{\tilde{q}_t}\Big)$, where $\tilde{q}_t$ is the 7-day Tukey-smoothed execution-demand benchmark & Execution + blob fee data; gas usage & \codepath{mart\_master\_daily.scarcity\_index} \\
$D^{*}_t$ & Latent demand factor & z-score & PC1 of standardized ETH log returns, CEX log volumes, realized volatility, Google Trends, and net stablecoin issuance (fit on pre-Dencun window; sign oriented so higher demand increases congestion) & Binance/OKX/Coinbase; Google Trends; issuer feeds & \codepath{demand\_factor\_daily.D\_star} \\
$\mathbf{R}_t$ & Regime dummies & binary & London, Merge, and post-Dencun indicators & Protocol calendar & \codepath{mart\_master\_daily.regime\_*} \\
$\mathbf{Cal}_t$ & Calendar dummies & binary & UTC weekend, month-end, quarter-turn indicators & Calendar & \codepath{mart\_master\_daily.calendar\_*} \\
$\mathbf{Shock}_t$ & Targeted events & binary & Event flags for airdrops, sequencer outages, mega NFT mints, market-stress days; catalog in Table~\ref{tab:targeted_events} & Curated event list & \codepath{controls\_shock\_daily.*} \\
\bottomrule
\end{tabular}
\end{table}

\subsection{Treatment Construction: Posting-Clean Adoption}
\label{sec:appendix:treatment_construction}
\begin{enumerate}[leftmargin=*]
    \item Pull daily L2 transaction counts by chain from \codepath{mart\_l2\_daily} and L1 user transactions from \codepath{stg\_l1\_blocks\_daily}.
    \item Identify L2$\rightarrow$L1 posting transactions via the rollup inbox registry (\codepath{l2\_inbox\_registry}); tag them in both datasets.
    \item Remove tagged posting transactions from the L2 numerator and the L1 denominator so that the treatment reflects end-user execution, not settlement load.
    \item Aggregate remaining L2 user transactions across tracked rollups (Arbitrum, Optimism, Base, zkSync, Starknet, Linea, Scroll) and compute $A^{clean}_t$ on the daily UTC grid; the full registry of inbox contracts and rollup identifiers lives in the replication bundle as \codepath{l2\_inbox\_registry}.
    \item Winsorize $A^{clean}_t$ at the 0.5\% tails and carry the resulting share into all confirmatory and exploratory designs.
\end{enumerate}

\subsection{Outcome Definitions and Units}
\begin{itemize}[leftmargin=*]
    \item \textbf{Base fee ($\log C^{fee}_t$).} Natural log of the median EIP-1559 base fee (Gwei) across blocks in day $t$.
    \item \textbf{Utilization ($u_t$).} Median block-level gas-used-to-gas-limit ratio per day, retaining the post-Merge 1.5 cap.
    \item \textbf{Scarcity index ($S_t$).} Combines execution gas and data-availability fees: daily median base fee + priority tip + (post-Dencun) blob base fee, divided by a 7-day smoothed demand benchmark $\tilde{q}_t$ (median gas used smoothed with a Tukey-Hanning window) and logged. This keeps scarcity comparable across London, Merge, and blob eras.
\end{itemize}

\subsection{Demand Factor $D^{*}_t$}
\begin{itemize}[leftmargin=*]
    \item \textbf{Inputs:} (i) ETH/USD log returns; (ii) log centralized-exchange spot volume (Binance, Coinbase, OKX aggregate); (iii) realized volatility from 5-minute returns; (iv) Google Trends ``ethereum'' index; (v) net stablecoin issuance (USDC + USDT + DAI).
    \item \textbf{Standardization and window:} Each series is z-scored using its mean and standard deviation over the London$\rightarrow$Merge window (2021-08-05 to 2024-03-12) to avoid blob-era structural breaks; single-day gaps are forward-filled before standardization.
    \item \textbf{PCA fit:} Principal components are estimated on the pre-Dencun standardized matrix; PC1 is rescaled to unit variance and sign-flipped so that higher $D^{*}_t$ aligns with higher fees.
    \item \textbf{Usage:} The same $D^{*}_t$ enters ITS, ECM, IV, and BSTS designs; sensitivity checks with ``lite'' inputs appear in Table~\ref{tab:i2_demand_factor}.
\end{itemize}

\subsection{Quality Control and Harmonization}
\begin{itemize}[leftmargin=*]
    \item \textbf{Time and aggregation.} All variables use UTC calendar days; block-level quantities are aggregated with medians to limit outlier influence.
    \item \textbf{Winsorization.} $A^{clean}_t$, $\log C^{fee}_t$, $u_t$, and $S_t$ are winsorized at the 0.5\% tails across the full sample ($N=1{,}245$) before entering regressions.
    \item \textbf{Missingness.} Days with missing treatment or base-fee fields ($<0.3\%$) are dropped listwise; PCA inputs with single-day gaps are forward-filled prior to z-scoring.
    \item \textbf{Smoothing choices.} The scarcity benchmark $\tilde{q}_t$ uses a 7-day Tukey-Hanning window; BSTS price baselines use daily mean and close prices as noted in Appendix~\ref{sec:appendix:bsts}.
\end{itemize}

\subsection{Targeted Shock Catalog}
\label{sec:appendix:shock_catalog}
\begin{small}
\begin{longtable}{p{0.10\textwidth} >{\raggedright\arraybackslash\seqsplit}p{0.21\textwidth} p{0.12\textwidth} p{0.12\textwidth} p{0.08\textwidth} >{\raggedright\arraybackslash}p{0.27\textwidth}}
\caption{Targeted Shock Catalog with Usage Flags}\label{tab:targeted_events}\\
\toprule
Category & Event & Date (UTC) & Used in confirmatory window? & Duration & Rationale \\
\midrule
\endfirsthead
\multicolumn{6}{l}{\footnotesize\textit{Table~\ref{tab:targeted_events} (continued)}}\\
\toprule
Category & Event & Date (UTC) & Used in confirmatory window? & Duration & Rationale \\
\midrule
\endhead
\midrule
\multicolumn{6}{r}{\footnotesize\textit{Continued on next page}}\\
\endfoot
\bottomrule
\endlastfoot
\multicolumn{6}{l}{\textit{Pre-Dencun (used in confirmatory window unless noted)}}\\
Protocol & London EIP-1559 & 2021-08-05 & Y & 1d & Fee-mechanism activation; sets baseline regime dummy. \\
Launch & Arbitrum One mainnet & 2021-09-01 & Y & 1d & Major L2 launch; sudden user migration. \\
Airdrop & dYdX airdrop & 2021-09-08 & Y & 1d & Large claim day; spikes L2+L1 usage. \\
Launch & Polygon Hermez v1 & 2021-03-01 & N & 1d & Pre-sample launch noted for completeness. \\
Airdrop & Immutable X airdrop & 2021-11-05 & Y & 1d & NFT airdrop; gas spike. \\
Launch & Starknet Alpha mainnet & 2021-11-16 & Y & 1d & Early Starknet deployment. \\
Launch & Optimism public mainnet & 2021-12-16 & Y & 1d & Public rollout; user onboarding burst. \\
Airdrop & Optimism airdrop 1 & 2022-05-31 & Y & 1d & First OP distribution; heavy claim traffic. \\
Upgrade & Arbitrum Nitro upgrade & 2022-08-31 & Y & 1d & Sequencer upgrade; throughput jump. \\
Protocol & Ethereum Merge & 2022-09-15 & Y & 1d & Consensus shift; volatility control. \\
Airdrop & Optimism airdrop 2 & 2023-02-09 & Y & 1d & Second OP claim event. \\
Airdrop & Arbitrum airdrop & 2023-03-23 & Y & 1d & ARB token claim; gas surge. \\
Launch & zkSync Era mainnet & 2023-03-24 & Y & 1d & zkSync Era launch. \\
Launch & Polygon zkEVM mainnet & 2023-03-27 & Y & 1d & Polygon zkEVM debut. \\
Upgrade & Optimism Bedrock upgrade & 2023-06-06 & Y & 1d & Bedrock migration; temporary pause/resume. \\
Launch & Linea mainnet & 2023-07-11 & Y & 1d & Linea mainnet go-live. \\
Launch & Mantle mainnet & 2023-07-17 & Y & 1d & Mantle mainnet go-live. \\
Campaign & Base Onchain Summer & 2023-08-09 & Y & 7d & Promo campaign; NFT mint surge. \\
Launch & Base mainnet & 2023-08-09 & Y & 1d & Base public launch. \\
Airdrop & Optimism airdrop 3 & 2023-09-18 & Y & 1d & Third OP claim wave. \\
Launch & Scroll mainnet & 2023-10-17 & Y & 1d & Scroll mainnet launch. \\
Campaign & Starknet STRK token launch & 2024-02-14 & Y & 1d & Token announcement; claim anticipation. \\
Airdrop & Optimism airdrop 4 & 2024-02-15 & Y & 1d & Fourth OP claim day. \\
Protocol & Dencun EIP-4844 & 2024-03-13 & N & 1d & Blob activation; start of exploratory blob era. \\
\midrule
\multicolumn{6}{l}{\textit{Post-Dencun (used in exploratory sensitivity only)}}\\
Airdrop & zkSync airdrop & 2024-06-17 & N & 1d & Large airdrop during blob era. \\
Upgrade & Polygon MATIC-to-POL transition & 2024-09-04 & N & 1d & Token transition; potential bridge congestion. \\
Campaign & Starknet staking launch & 2024-11-26 & N & 1d & Staking launch; sequencer load risk. \\
\end{longtable}
\end{small}
\vspace{-0.5em}
\begin{minipage}{0.95\textwidth}
\footnotesize\textit{Note:} Column~4 flags inclusion in the confirmatory London$\rightarrow$Dencun window; post-Dencun events are retained for exploratory robustness only. Duration records the anchor day used in regressions (multi-day campaigns are coded with a single start-day dummy). Rationale summarizes why the event could jointly shift adoption and congestion.
\end{minipage}

\end{document}